\documentclass[a4paper,11pt]{article}

\usepackage{amsmath}
\usepackage{amssymb}

\usepackage[colorlinks=true, allcolors=blue]{hyperref}
\usepackage{graphicx}% Include figure files
\usepackage{dcolumn}% Align table columns on decimal point
\usepackage{bm}% bold math
\usepackage[dvipsnames]{xcolor}

\newcommand{\rc}[1]{\textcolor{black}{#1}}

\newcommand{\dg}[1]{${#1}^{\circ}$}
\usepackage{fancyhdr}
\usepackage[titletoc]{appendix}
\usepackage{multicol}
\usepackage{subcaption}
\usepackage{bm}
\usepackage{amsfonts}
\usepackage{epstopdf}
\usepackage{xcolor}
\usepackage{color}
\usepackage{wasysym}
\usepackage{hhline}
\usepackage[capitalise]{cleveref}
\usepackage{authblk}
\usepackage{upgreek}
\newcommand{\micron}{{\upmu\mathrm{m}}}

\begin{document}

%\preprint{APS/123-QED}

\title{\rc{Electron Acceleration Using Twisted Laser Wavefronts}}

\author[1,2]{Yin Shi}
\author[2]{David Blackman}
\author[2]{Alexey Arefiev}%
\affil[1]{Department of Plasma Physics and Fusion Engineering, University of Science and Technology of China, Hefei 230026, China}%
\affil[2]{Department of Mechanical and Aerospace Engineering, University of California at San Diego, La Jolla, CA 92093, USA}%

\date{\today}% It is always \today, today,
  % but any date may be explicitly specified
\maketitle

\begin{abstract}
{Using plasma mirror injection we demonstrate, both analytically and numerically, that a circularly polarized helical laser pulse can accelerate highly collimated dense bunches of electrons to several hundred MeV using currently available laser systems. The circular-polarized helical (Laguerre-Gaussian) beam has a unique field structure where the transverse fields have helix-like wave-fronts which tend to zero on-axis where, at focus, there are large on-axis longitudinal magnetic and electric fields. The acceleration of electrons by this type of laser pulse is analysed as a function of radial mode number and it is shown that the radial mode number has a profound effect on electron acceleration close to the laser axis.
\rc{Using three-dimensional particle-in-cell simulations a circular-polarized helical laser beam with power of 0.6~PW is shown to produce several dense attosecond bunches. The bunch nearest the peak of the laser envelope has an energy of 0.47~GeV with spread as narrow as 10\%, a charge of 26~pC with  duration of $\sim 400$~as, and a very low divergence of 20~mrad}. The confinement by longitudinal magnetic fields in the near-axis region allows the longitudinal electric fields to accelerate the electrons over a long period after the initial reflection. Both the longitudinal E and B fields are shown to be essential for electron acceleration in this scheme. This opens up new paths towards attosecond electron beams, or attosecond radiation, at many laser facilities around the world.}

\end{abstract}

\section{Introduction}\label{Sec-1}

Laser-driven electron accelerators have become a very active area of research due to technological developments and improvements of high-power laser beams~\cite{Danson2019}. There are usually two common approaches. One approach is laser wakefield acceleration~\cite{Esarey2009} that utilizes plasma electric fields whose strength is related to the plasma density. Another approach is direct laser acceleration~\cite{gibbon2004short} that relies on the fields of the laser for the energy transfer inside a plasma (e.g. see \cite{Arefiev_POP_2016}) or in vacuum~\cite{Stupakov_PRL_2001}. In the vacuum regime, transverse electron expulsion typically terminates the energy gain and leads to strong electron divergence. This is the reason why direct laser acceleration in vacuum has been deemed ineffective and most of the research has been focused on the plasma regime where collective fields prevent the expulsion~\cite{Pukhov_1999, Willingale_PRL_2011,gong.pre.2020}. In an attempt to mitigate the expulsion in the vacuum regime, some studies examined alternative approaches utilizing longitudinal electric fields of a radially polarized beam~\cite{Zaim2017} and higher-order Gaussian beams~\cite{Sprangle1996} for electron acceleration.

Currently there are several multi-PW laser systems operational in the world, with several more due to come online in the next few years~\cite{Danson2019}. The biggest of the new systems recently proposed, at the Shanghai Superintense-Ultrafast Laser Facility~\cite{Shen2018}, is pushing the 100~PW limit and will be in development over the next decade. At the same time, new optical techniques using helical wave-fronts~\cite{Leblanc2017, shi2014} are being developed. One of the key advantages of the helical, or Laguerre-Gaussian, laser mode is that it can be produced, at high efficiency, from a standard Gaussian laser pulse in reflection from a fan-like structure~\cite{shi2014,Longman2017}. Very significant differences can be seen when comparing the laser-plasma interactions of conventional laser beams to those of helical beams, some of which have been examined in simulations~\cite{vieira2016, Zhang2015, vieira2018, Shi2018,Longman2017, Ju2018, Zhu2019,TIKHONCHUK2020,Nuter2018,Blackman2020}, and some have begun to be explored in recent experiments~\cite{Leblanc2017,Denoeud2017,Longman2020, BAE2020, Aboushelbaya2020}. An electron acceleration scheme has recently been proposed where a high-power high-intensity circular-polarized Laguerre-Gaussian beam is reflected from a plasma mirror~\cite{Shi2021}. The unique field structure of this beam both confines and accelerates tightly packed electron bunches to GeV energies with \rc{a narrow energy spread}.

In this article, we show using 3D particle-in-cell (PIC) simulations that the same electron acceleration scheme can be successfully applied 1) using significantly lower laser power than that used in~\cite{Shi2021} and 2) using oblique incidence. Specifically, we show that a 600~TW laser beam generates \rc{several dense attosecond bunches at intervals similar to the laser wavelength. The bunch nearest the peak of the laser envelope gains an energy of 0.47~GeV (10\% FWHM energy spread), while maintaining a bunch charge as high as 26~pC. } The tightly confined bunch has a duration of $\sim 400$~as and a remarkably low divergence of just \dg{1.15} (20~mrad). In addition to this, the scheme is demonstrated to be tenable with an angle of incidence as high as \dg{25}. Our results show that the electron acceleration by helical beams is not limited to high-power high-intensity lasers and can be successfully explored at a wide range of laser facilities.

The rest of this paper is organized as follows. \Cref{Sec-2} examines the field structure of linearly and circularly polarized laser beams with twisted \rc{wavefronts}. The near-axis structure of the longitudinal electric field for different radial modes is explicitly derived. \Cref{sec:estimates} is dedicated to estimating the acceleration that the longitudinal fields can provide and the effect of the radial mode structure of the laser on this acceleration. In Section~\ref{Sec-sim}, we present results of a 3D PIC simulation for a 600~TW laser beam with twisted \rc{wavefronts} whose near-axis field structure is dominated by longitudinal electric and magnetic fields. Section~\ref{Sec-bx} is concerned with exploring the role of the longitudinal magnetic field where the effect of this field on the bunch density and particle trajectories is shown, including the possible reliability of this model with an oblique angle of incidence. In Section~\ref{Sec-6}, we summarise the main results of this work.

%++++++++++++++++++++++++++++++++++++++++++++++

\section{Field topology of a laser beam with twisted wavefronts} \label{Sec-2} 

In this section, we examine the field topology of linearly and circularly polarized laser beams with twisted \rc{wavefronts}. We show that, for a properly chosen twist index, the field structure in the region close to the axis of the beam can primarily consist of longitudinal electric and magnetic fields. The differences between radial modes are explicitly emphasized. 

\subsection{Linearly polarized beam}

We start by considering a linearly polarized laser beam with wavelength $\lambda_0$ propagating in vacuum along the $x$-axis. We assume that the diffraction angle, defined as $\theta_d = w_0/x_R$, is small, where $w_0$ is the beam waist, and $x_R = \pi w_0^2 / \lambda_0$ is the Rayleigh range. Without any loss of generality, we assume that the laser electric field is polarized along the $y$-axis. In this case, it is convenient to describe the field structure using a vector potential $\bm{A}$ that satisfies the Lorenz gauge condition and has only one non-zero component, $A_y$. The paraxial wave equation for $A_y$ has the form
\begin{equation} \label{paraxial_equation} 
 \left[ \frac{\partial^2}{\partial \widetilde{y}^2} + \frac{\partial^2}{\partial \widetilde{z}^2} + 4i \frac{\partial}{\partial \widetilde{x}} \right] A_y= 0,
\end{equation}
where, for compactness, the longitudinal coordinate $x$ is normalized to $x_R$ and the transverse coordinates $y$ and $z$ are normalized to $w_0$:
\begin{eqnarray}
    && \widetilde{x} = x/x_R, \\
    && \widetilde{y} = y/w_0, \\
    && \widetilde{z} = z/w_0. 
\end{eqnarray}
\Cref{paraxial_equation} implies the following form of the solution with $\omega = 2 \pi c / \lambda_0$: 
\begin{eqnarray}
 && A_y = \Psi_y (\widetilde{x},\widetilde{y},\widetilde{z}) g(\xi) \exp(i\xi),
\end{eqnarray}
where $g$ is the envelope function with $\max (g) = 1$ and 
\begin{equation}
    \xi \equiv 2\widetilde{x}/\theta_d^2 - \omega t
\end{equation}
is the phase variable. 

The transverse electric field in the paraxial approximation, i.e. at $\theta_d \ll 1$, is given by
\begin{equation} \label{eq:E_y}
 E_y \approx - \frac{1}{c} \frac{\partial A_y}{\partial t} = \frac{i \omega}{c} A_y.
\end{equation}
We are interested in solutions {of the form}
\begin{equation}
    E_y = E_{0}  g(\xi) \exp(i \xi) \psi_{l,p} (\widetilde{x},\widetilde{r},\phi) \label{E_y_1}
\end{equation}
where  
\begin{equation} \label{eq:psi}
    \psi_{l,p} (\widetilde{x},\widetilde{r},\phi) = C_{p, l} f(\widetilde{x})^{|l| + 1 + 2p} (1+\widetilde{x}^2)^p  L_{p}^{|l|}\left(\frac{2\widetilde{r}^2}{1+\widetilde{x}^2}\right) \left(\sqrt{2}\widetilde{r}\right)^{|l|} \exp \left[ -\widetilde{r}^2 f(\widetilde{x})\right] \exp\left(i l \phi\right) 
\end{equation}
is a mode with a twist index $l$ and radial index $p$. Here we introduced

\begin{equation}
 \widetilde{r} = \sqrt{ \widetilde{y}^2 + \widetilde{z}^2}, \quad  \phi = \arctan ( \widetilde{z}/ \widetilde{y}), \quad  f(\widetilde{x}) = \frac{1-i\widetilde{x}}{1+\widetilde{x}^2}. \label{E_y_3}
\end{equation}

The $L_{p}^{|l|}$ function is the generalized Laguerre polynomial and $C_{p, l}$ is a normalization constant. The modes $\psi_{l,p}(\widetilde{x},\widetilde{r},\phi)$ are orthonormal at a given  $\widetilde{x}$~\cite{Allen1992}, with
\begin{equation}
    C_{p,l} = \sqrt{\frac{2p!}{\pi (p + |l|)!}},
\end{equation}
such that
\begin{equation} \label{eq:norm}
    \int_0^{2\pi} d \phi \int_{0}^{\infty} \psi_{l,p}(\widetilde{x},\widetilde{r},\phi) \psi^*_{l,p}(\widetilde{x},\widetilde{r},\phi) \widetilde{r} d \widetilde{r} = 1.
\end{equation}
It is worth pointing out that $E_0$ in \cref{E_y_1} is not the peak amplitude of $E_y$. It is therefore convenient to introduce 
\begin{equation}
    E_0^* \equiv \sqrt{\frac{2}{\pi}} E_0,
\end{equation}
which is the peak amplitude of the transverse electric field in the beam with $l = 0$.

The transverse magnetic field has a $z$-component only, since $\bm{B} = \nabla \times \bm{A}$. In the paraxial approximation, we have
\begin{equation}
 B_z = \frac{1}{x_R} \frac{\partial A_y}{\partial \widetilde{x}} \approx \frac{2 i }{\theta_d w_0} A_y.
\end{equation}
After taking into account that $\theta_d = \lambda_0 / \pi w_0$, we find that $B_z = E_y$. The longitudinal component of the Poynting vector $\bm{S}$ is then given by $S_x = (c/4 \pi) \left( \mbox{Re} E_y \right)^2$ and the peak period-averaged power for the linearly polarized beam is
\begin{equation} \label{eq:power_poynting_int}
    P_{lin} = \left\langle \int_0^{2\pi} d \phi \int_{0}^{\infty} S_x r d r \right\rangle = \frac{c w_0^2}{4 \pi}  \int_0^{2\pi} d \phi \int_{0}^{\infty} \left\langle \left( \mbox{Re} E_y \right)^2 \right\rangle \widetilde{r} d \widetilde{r} ,
\end{equation}
where the angle brackets indicate the time-averaging. It is convenient to compute the power in the focal plane, i.e. at $\widetilde{x} = 0$. We use the expression given by \cref{E_y_1} to find that
\begin{equation} \label{eq:power}
    P_{lin} = \frac{c w_0^2}{8 \pi} E_0^2 = \frac{\pi}{2} \frac{a_0^2 w_0^2 m_e^2  c^5}{\lambda_0^2 e^2},
\end{equation}
where
\begin{equation}
    a_0 = |e|E_0/m_e c \omega
\end{equation}
is a dimensionless parameter, and $e$ and $m_e$ are the electron charge and mass. We have explicitly taken into account the normalization condition given by \cref{eq:norm}. The advantage of the chosen normalization is that the power $P$ is the same for different modes with the same $E_0$ or $a_0$.

The longitudinal laser electric and magnetic fields can be calculated from the $(\nabla \cdot \bm{E}) = 0$ and $(\nabla \cdot \bm{B}) = 0$ conditions, respectively. In the paraxial approximation, we have
\begin{eqnarray}
 E_x \approx \frac{i \theta_d}{2} \frac{\partial E_y}{\partial \widetilde{y}}, &\mbox{ }& B_x \approx \frac{i \theta_d}{2} \frac{\partial E_y}{\partial \widetilde{z}}.
\end{eqnarray}
It follows from Eqs.~(\ref{E_y_1})~-~(\ref{E_y_3}) that
\begin{eqnarray}
 && E_x^{\pm} = \begin{cases}
 \frac{i \theta_d}{2} \left[ \frac{|l|}{\widetilde{r}} e^{\mp i \phi} - 2f\widetilde{r} \cos \phi - \frac{2}{1+\widetilde{x}^2} \frac{L_{p - 1}^{|l| +1}}{L_{p}^{|l|}}\cos \phi \right] E_y ; \mbox{   for } p \geq 1 , \\
 \frac{i \theta_d}{2} \left[ \frac{|l|}{\widetilde{r}} e^{\mp i \phi} - 2f\widetilde{r} \cos \phi \right] E_y ; \mbox{   for } p = 0 , \label{E_z_pm1}
 \end{cases} \\
 && B_x^{\pm} = \begin{cases}
 \frac{\theta_d}{2} \left[ \mp \frac{|l|}{\widetilde{r}} e^{\mp i \phi} - 2i f\widetilde{r} \sin \phi - \frac{2i}{1+\widetilde{x}^2} \frac{L_{p - 1}^{|l| +1}}{L_{p}^{|l|}}\sin \phi \right] E_y ; \mbox{   for } p \geq 1 , \\
 \frac{\theta_d}{2} \left[ \mp \frac{|l|}{\widetilde{r}} e^{\mp i \phi} - 2i f\widetilde{r} \sin \phi \right] E_y ; \mbox{   for } p = 0 ,\label{B_z_pm1}
 \end{cases} 
\end{eqnarray}
where the superscript on the left-hand side represents the sign of $l$. 

The twist of the field represented by $l$ qualitatively changes the topology of the transverse and longitudinal fields. We are particularly interested in the field structure close to the central axis, i.e. at $\widetilde{r} \rightarrow 0$. It is important to distinguish three cases based on the value of the twist index: $l = 0$, $|l| = 1$, and $|l| > 1$. In the near-axis region, we have $E_y \propto \widetilde{r}^{|l|} \exp(i l \phi)$. In the case of $l = 0$ or a beam without a twist, the longitudinal fields vanish on the central axis, while the transverse fields reach their maximum value. In the case of $|l| > 1$, all laser fields vanish on the central axis. The most unusual is the case of $|l| = 1$, because in this case the longitudinal rather than transverse fields peak on axis. As a result, the near-axis field structure is dominated by longitudinal fields. Note that according to Eqs.~(\ref{E_z_pm1}) and (\ref{B_z_pm1}) these fields are not axis-symmetric.

\begin{figure}
 \centering
 \includegraphics[width=0.49\linewidth]{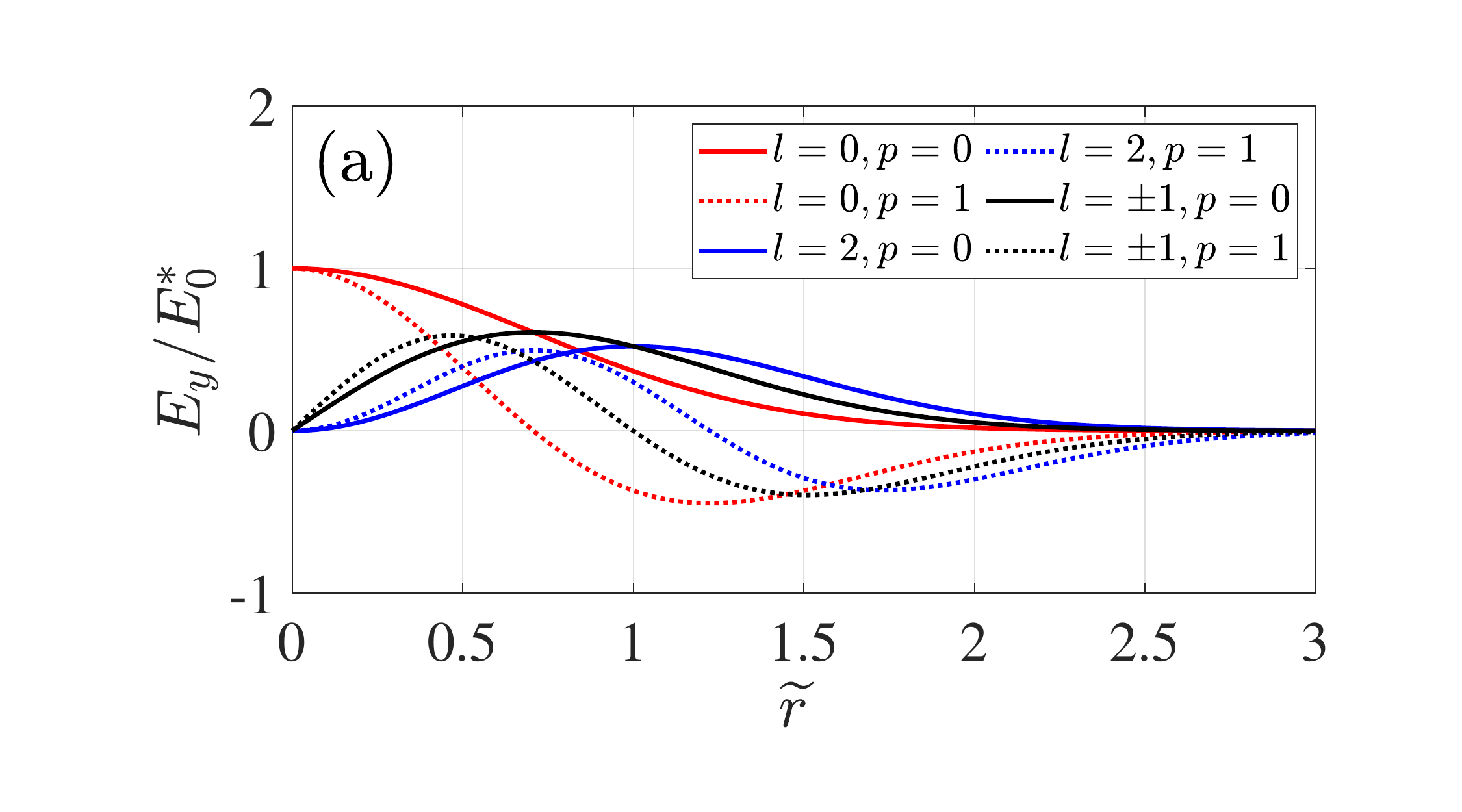} %LG-CP Detailed paper/Exyz_discuss_a1.pdf
 \includegraphics[width=0.49\linewidth]{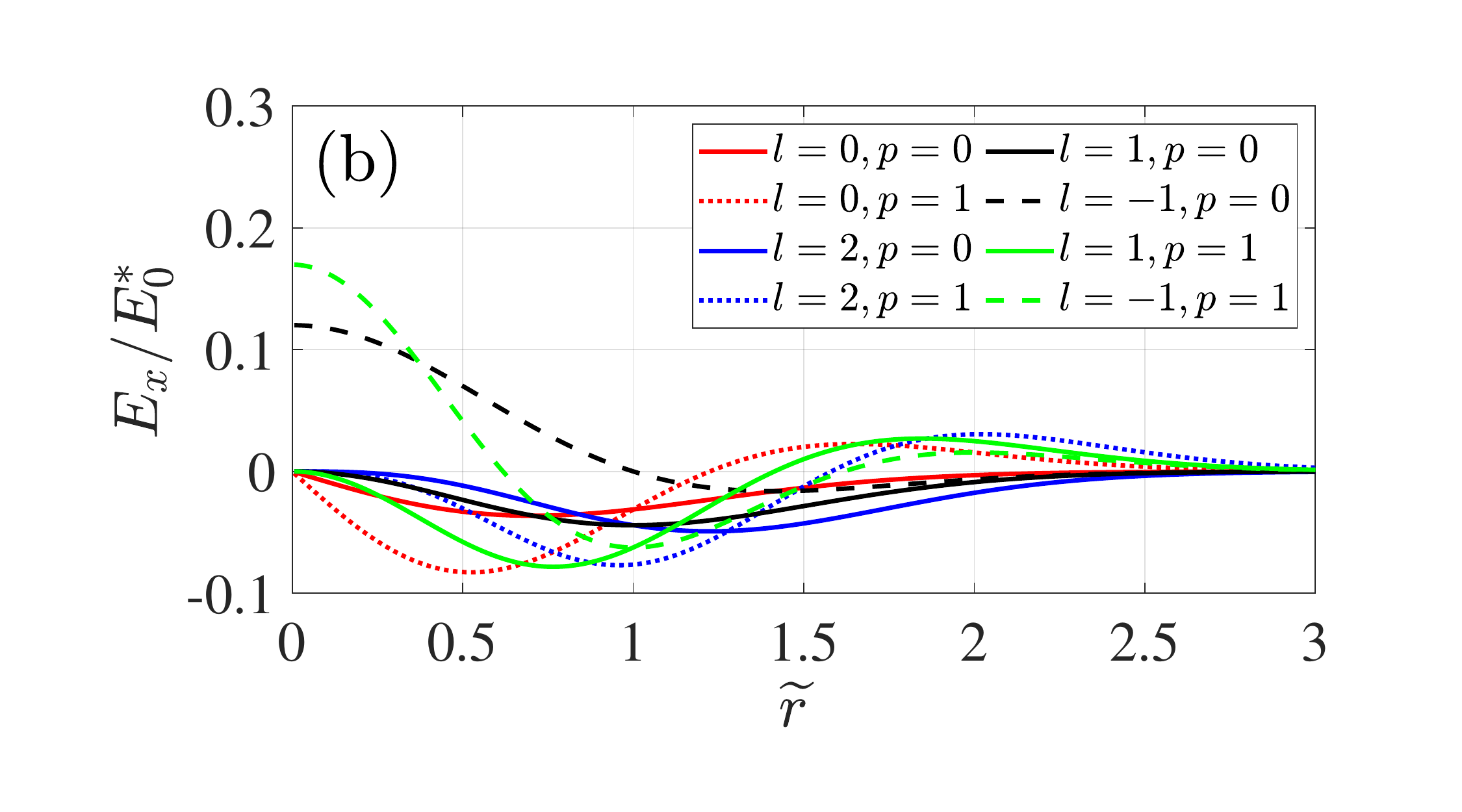} %LG-CP Detailed paper/Exyz_discuss_b1.pdf
\caption{Transverse and longitudinal electric field structure in the focal plane, $\widetilde{x} = 0$, of right-circularly polarized beams ($\sigma = 1$) with $l = 0, \pm 1$, 2 and $p = $0, 1. All beams have the same $E_0$ and the same diffraction angle, $\theta_d = 8.5 \times 10^{-2}$. The fields are normalized to $E_0^{*}$, which is the peak amplitude of the transverse electric field in the beam with $l = 0$.} \label{Exyz_discuss1}
\end{figure}

\subsection{Circularly polarized beam}

The results of the previous subsection can be readily generalized to the case of a circularly polarized laser pulse. In addition to $E_y$, the laser beam also has an $E_z$-component. We set $E_z = i \sigma E_y$, where $\sigma = 1$ corresponds to the right-circularly polarized wave and $\sigma = -1$ corresponds to the left-circularly polarized wave. The longitudinal electric and magnetic fields can again be calculated using the $(\nabla \cdot \bm{E}) = 0$ and $(\nabla \cdot \bm{B}) = 0$ equations, respectively. In the paraxial approximation, we have
\begin{eqnarray}
 && E_x \approx \frac{i \theta_d}{2} \left( \frac{\partial E_y}{\partial \widetilde{y}} + \frac{\partial E_z}{\partial \widetilde{z}} \right), \\
 && B_x \approx \frac{i \theta_d}{2} \left( \frac{\partial E_y}{\partial \widetilde{z}} - \frac{\partial E_z}{\partial \widetilde{y}} \right).
\end{eqnarray}
After substituting $E_z = i \sigma E_y$ into these equations, we find that
\begin{eqnarray}
 && E_x^{\pm} = \begin{cases}
 i \theta_d E_y \left[ \frac{|l|}{\widetilde{r}} \frac{1 \mp \sigma }{2} e^{\mp i \phi} - \widetilde{r} f e^{ i \sigma \phi} - \frac{1}{1+\widetilde{x}^2} \frac{L_{p - 1}^{|l| +1}}{L_{p}^{|l|}}e^{ i \sigma \phi} \right]; \mbox{   } p \geq 1 , \\
 i \theta_d E_y \left[ \frac{|l|}{\widetilde{r}} \frac{1 \mp \sigma }{2} e^{\mp i \phi} - \widetilde{r} f e^{ i \sigma \phi} \right]; \mbox{   } p = 0 , \label{E_z_pm-C}
 \end{cases} \\
 && B_x^{\pm} = \begin{cases}
 \theta_d E_y \left[ \frac{|l|}{\widetilde{r}} \frac{\sigma \mp 1}{2} e^{\mp i \phi} - \widetilde{r} f \sigma e^{ i \sigma \phi} - \frac{\sigma}{1+\widetilde{x}^2} \frac{L_{p - 1}^{|l| +1}}{L_{p}^{|l|}}e^{ i \sigma \phi} \right] ; \mbox{   } p \geq 1 , \\
 \theta_d E_y \left[ \frac{|l|}{\widetilde{r}} \frac{\sigma \mp 1}{2} e^{\mp i \phi} - \widetilde{r} f \sigma e^{ i \sigma \phi} \right] ; \mbox{   } p = 0 ,\label{B_z_pm-C}
 \end{cases} 
\end{eqnarray}
where the superscripts on the left-hand side again represent the sign of $l$. 

Only for $|l| = 1$ the longitudinal rather than transverse fields peak on axis. However, the two circular polarizations (right and left) are not equivalent. In the case of $\sigma = - l$, the longitudinal fields reach their highest amplitude at $\widetilde{r} \rightarrow 0$. On the other hand, in the case of $\sigma = l$, the longitudinal fields vanishes on axis. The transverse fields vanish on axis in both cases. It is worth pointing out that, in contrast to the linearly polarized beam, the longitudinal fields of the circularly polarized beam with $\sigma = - l$ are axis-symmetric. 

\Cref{Exyz_discuss1} shows the field structure for right-circularly polarized beams ($\sigma = 1$) with different values of the twist index $l$ and radial index $p$. In agreement with our analysis, the transverse field in \cref{Exyz_discuss1}(a) only peaks on axis for the beams without a twist ($l = 0$). In all other cases, the transverse field vanishes at $\widetilde{r} \rightarrow 0$. The longitudinal field, shown in \cref{Exyz_discuss1}(b), only peaks on axis for the beam with $l = -1$. As predicted, the longitudinal field vanishes at $\widetilde{r} \rightarrow 0$ for $l = 1$.

\begin{figure}
    \centering
    \includegraphics[width=1\textwidth]{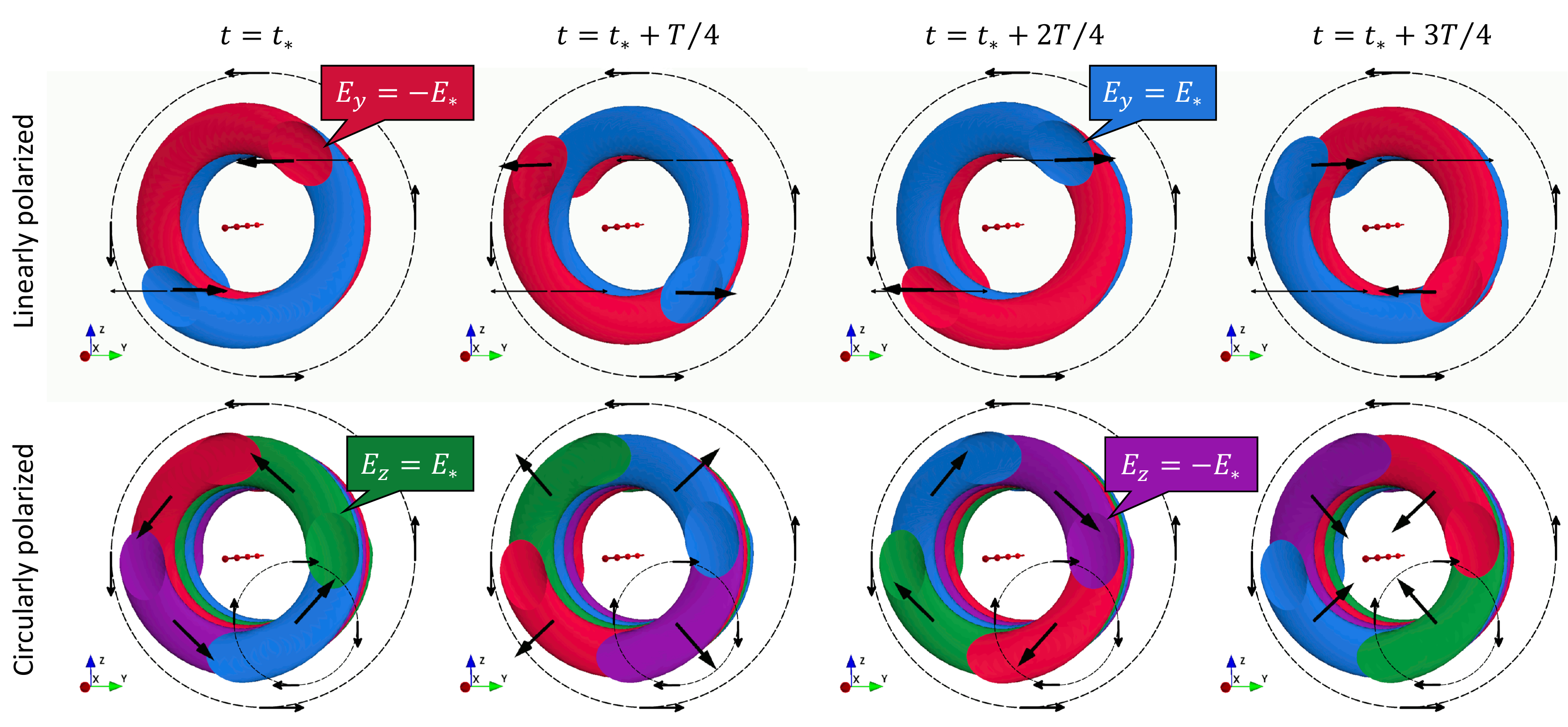}
    \caption{Schematic representation of the transverse electric field topology for a linearly polarized beam (top row) with $l=-1$ and $p=0$ and a circularly polarized beam (bottom row) with $l=-1$, $p=0$, and $\sigma=-l$. The diagrams show one wavelength near the focal plane of each beam. The color-coded surfaces are the surfaces of the constant amplitude $|\bm{E}_\perp| = E_*$: blue is  positive $E_y$, red is negative $E_y$, green is positive $E_z$, and purple is negative $E_z$.
    The large arrows show the direction of $\bm{E}_\perp$ at a given location. The small black arrows placed on the large circles represent the motion of $|\bm{E}_\perp|$ peaks over time. The small black arrows placed on the small circles in the bottom row represent the direction of rotation of the electric field over time. The red structure at the centre of each image represents the the central axis of each beam. The images from left to right are at times incremented by a quarter period $T/4$, where $T = 2 \pi/\omega$. }
    \label{fig:rotation_schematic}
\end{figure}

It must be pointed out that the circularly polarized beams with $|l| = 1$ have two, rather than one, rotations to consider: the rotation of the electric field maxima about the central axis due to the \rc{wavefronts} twist and the rotation of the electric field vector due to the choice of polarization. In the case with $|l| = 1$ and $\sigma = - l$, the rotation of the transverse electric field vector due to the polarization and the twist of the transverse field \rc{wavefronts} have opposing chiralities. \Cref{fig:rotation_schematic} (bottom row)  provides schematic diagrams showing the two  rotations. For comparison, the upper row in \cref{fig:rotation_schematic}) shows the field topology for a linearly polarized beam, with $E_\perp = E_y$. Recall that the circularly polarized beam is a superposition of two linearly polarized beams.

In the sections that follow, we focus on electron acceleration in the near-axis region by $E_x$ of a circularly polarized laser beam with $l = -1$ and $\sigma = 1$. In order to obtain a compact expression for $E_x$, we set $\widetilde{r} = 0$ in Eq.~(\ref{E_z_pm-C}) and substitute the expression for $E_y$ given by Eq.~(\ref{E_y_1}), which yields
\begin{equation} \label{Eparallel}
 E_{x}{(\widetilde{r}=0)} = iE_{\parallel}^{\max} f^{2 (1 + p)}( 1+ \widetilde{x}^2)^p g( \xi ) e^{ i \xi } = \frac{i g( \xi ) E_{\parallel}^{\max}}{1+ \widetilde{x}^2} \exp \left[ i \xi - i 2(1+p) \tan^{-1} \widetilde{x} \right]  ,
\end{equation}
where 
\begin{equation} \label{eq:E_max}
    E_{\parallel}^{\max} = 2 \sqrt{\frac{p +1}{\pi}} \theta_d E_0 = \frac{2}{\pi} \sqrt{\frac{p +1}{\pi}} \frac{\lambda_0}{w_0} E_0
\end{equation}
is the peak value of the longitudinal electric field. A feature that is important for electron acceleration is the explicit dependence of the phase on $p$. This will be discussed in more detail in \cref{sec:estimates}. The peak period-averaged power $P$ for the circularly-polarized beam is twice the power $P_{lin}$ for a single linearly polarized beam with the same amplitude $E_0$. Taking this into account and using the expression for $P_{lin}$ given by \cref{eq:power}, we obtain the following expression for $P$ in terms of the amplitude of the longitudinal field:
\begin{equation} \label{eq:power_circ}
 P =  \frac{\pi^4}{4(p+1)} \frac{w_0^4}{\lambda_0^4} \frac{m_e^2 c^5}{e^2} a_{\parallel}^2,
\end{equation}
where $a_{\parallel} \equiv |e| E_{\parallel}^{\max} / m_e c \omega$. \Cref{eq:power_circ} can be recast as an expression for normalized amplitudes of longitudinal electric and magnetic fields for a given power $P$ in the laser beam: 
\begin{equation} \label{a and P}
 a_{\parallel} \equiv \frac{|e| E_{\parallel}^{\max}}{m_e c \omega} = \frac{|e| B_{\parallel}^{\max}}{m_e c \omega} \approx 71\sqrt{p +1}\left( \frac{\lambda_0}{w_0} \right)^2 P^{1/2} [\mbox{PW}].
\end{equation}

We conclude this section by pointing out that the longitudinal fields can be strong even for $\theta_d \ll 1$. Let us take a $P = 0.6$~PW circularly-polarized beam with $l = -1$, $\sigma = 1$, and $p = 0$. The focal spot size is $w_0 = 3.0~\micron$ and the wavelength is $\lambda_0 = 0.8~\micron$. In this case, $\theta_d \approx 8.5 \times 10^{-2}$, so our analysis that was performed in the paraxial approximation is applicable to this beam. It follows from \cref{eq:E_max} that $E_{\parallel}^{\max} \approx 0.1 E_0$.
Taking into account that {$E_{y}^{\max} = E_0 C_{0, l}|l|^{|l|/2}\exp{(-|l|/2)}$ for $p=0$}, we find that $E_{\parallel}^{\max} /E_{y}^{\max} = B_{\parallel}^{\max} /B_{y}^{\max} \approx 0.2$. We also find from \cref{a and P} that $a_{\parallel} \approx 3.8$. The corresponding dimensional field amplitudes are $E_{\parallel}^{\max} \approx 1.5 \times 10^{13}$~V/m and $B_{\parallel}^{\max} \approx 51$~kT. 

%++++++++++++++++++++++++++++++++++++++++++++++
%++++++++++++++++++++++++++++++++++++++++++++++

\section{Preliminary estimates for electron acceleration} \label{sec:estimates}

In this section, we perform preliminary estimates for electrons accelerated in the near-axis region by a helical laser beam whose field structure in this region predominantly consists of longitudinal electric and magnetic fields. We assume that the electrons are injected into the laser beam near the focal plane located at $\widetilde{x} = 0$. The injection is implied to occur when an incident beam is reflected off a mirror at normal incidence. This process is examined self-consistently in \cref{Sec-sim} and \cref{Sec-bx} using kinetic simulations. 

The phase velocity, $v_{ph}$, of $E_\parallel$ \rc{wavefronts} plays a key role in electron acceleration along the central axis. For simplicity, we limit our analysis to the part of the pulse that is near the peak of the envelope, which means that we can set $g(\xi) \approx 1$ in \cref{Eparallel}. In order to determine $v_{ph}$, it is convenient to re-write the expression for the longitudinal electric field given by \cref{Eparallel} as
\begin{equation}
    E_{\parallel} = - \frac{E_{\parallel}^{\max} \sin \left( \Phi + \Phi_0 \right)}{1 + x^2/x_R^2}, \label{E_||}
\end{equation} 
where the phase $\Phi$ is given by
\begin{equation}
    \Phi = 2 \left[ \theta_d^{-2} (x/x_R) - (p +1)\tan^{-1} (x/x_R) \right] - \omega t. \label{phase}
\end{equation}
The constant $\Phi_0$ can be interpreted as the injection phase for an electron that starts its acceleration at $x \approx 0$ at $t \approx 0$. We define the phase velocity as $v_{ph} =  dx/dt$ for $\Phi = \mbox{const}$. We differentiate \cref{phase}, where $\Phi = \mbox{const}$, to obtain 
\begin{equation}\label{eq:v_phase}
    0 = \frac{v_{ph}}{c} - \frac{v_{ph}}{c} \frac{(p + 1) \theta_d^2}{1 + x^2/x^2_R}  - 1,
\end{equation}
where it was taken into account that $\theta_d^2 = \lambda_0 / \pi x_R$.
In the paraxial approximation ($\theta_d \ll 1$ ), the second term on the right-hand side is small. We neglect this term to find that $v_{ph}/c \approx 1$. In order to find the correction associated with $\theta_d$, we set $v_{ph}/c = 1$ in the second term on the right-hand side of \cref{eq:v_phase} and obtain the following expression for the relative degree of superluminosity along the central axis:
\begin{equation} \label{eq:superlum}
    \frac{v_{ph} - c}{c} \approx \frac{(p + 1) \theta_d^2}{1 + x^2/x^2_R}.
\end{equation}
The key feature here is the explicit dependence on the radial index $p$, with the superluminosity being higher for higher-order radial modes.

The electron unavoidably slips with respect to the \rc{wavefronts} as it moves forward, which limits its energy gain. The slipping is determined by the difference $v_{ph} - v_x$, where $v_{ph} \geq c$ and $v_x < c$. As the electron becomes ultra-relativistic due to the acceleration by $E_\parallel$, it enters a regime where $c - v_x \ll v_{ph} - c$. In this regime, the slipping, or dephasing, is primarily determined by the relative degree of superluminosity given by \cref{eq:superlum}. The lowest estimate for the phase slip experienced by an electron that has travelled from $x_0$ to $x$ is obtained by setting $x = x_0 + c (t-t_0)$ in \cref{phase}, which yields
\begin{equation} \label{eq:Delta_Phi}
    \Delta \Phi = 2(p +1) \left[ \tan^{-1} (x_0/x_R) - \tan^{-1} (x/x_R) \right]
\end{equation}
For $x_0 \ll x_R$, this expression reduces to $\Delta \Phi = -2(p +1) \tan^{-1} (x/x_R)$. The phase velocity is superluminal near $x = 0$, but it decreases to $c$ at $x \gg x_R$. As a result, the total phase slip is finite and it approaches $\Delta \Phi = - (p +1)\pi$ at $x \gg x_R$. There is a significant difference between the $p=0$ mode and higher order radial modes. At $p=0$, the phase slip is $\Delta \Phi = - \pi$, which means that some electrons (this depends on the initial phase) can remain in the accelerating phase of $E_\parallel$ until the laser defocuses ($x \gg x_R$) and $E_\parallel$ becomes very weak. In contrast to that, all electrons experience deceleration by $E_\parallel$ at $p \geq 1$, because they slip into the decelerating phase prior to strong defocusing at $x \gg x_R$.

In order to estimate the electron energy gain from $E_\parallel$, we assume that the electron is ultra-relativistic with $c - v_x \ll v_{ph} - c$. In this case, the phase $\Phi$ in \cref{E_||} can be replaced by $\Phi \approx \Phi_0 + \Delta \Phi$, where $\Delta \Phi$ is given by the already derived \cref{eq:Delta_Phi}. The change in electron momentum during the acceleration is obtained by integrating the momentum balance equation $dp_\parallel / dt = -|e| E_\parallel$, which yields
\begin{equation}
 \Delta p_{\parallel} = |e| E_{\parallel}^{\max} \int_{t_0}^{t}  \frac{ \sin (\Delta \Phi + \Phi_0 + \pi) dt'}{1 + (x')^2 / x_R^2} ,
\end{equation} 
where $x'$ is the electron location at time $t'$. We note that $dx' / dt' \approx c$ in the considered regime, so we can switch from integration over time to integration over the longitudinal coordinate by replacing $dt'$ with $dx' / c$. After substituting the expression for \cref{eq:Delta_Phi}, we obtain
\begin{eqnarray} 
 \Delta p_{\parallel} &=& \frac{|e| E_{\parallel}^{\max}}{c} \int_{x_0}^{x}  \frac{ \sin (\Delta \Phi + \Phi_0) dx'}{1 + (x')^2 / x_R^2} \nonumber \\
 &=& -\frac{|e| E_{\parallel}^{\max} x_R}{2(p+1)c} \left( \cos \Phi_0 - \cos \left[ \Phi_0 + 2(p+1) \tan^{-1} (x_0/x_R) - 2(p+1)\tan^{-1} (x/x_R) \right] \right). 
\end{eqnarray} 
This result can be further simplified by assuming that $x_0 \ll x_R$, so that $\tan^{-1} (x_0/x_R) \approx 0$ and
\begin{equation}
    \Delta p_\parallel = -\frac{|e| E_{\parallel}^{\max} x_R}{2(p+1)c} \left( \cos \Phi_0 - \cos \left[ \Phi_0  - 2(p+1)\tan^{-1} (x/x_R) \right] \right). \label{main_result}
\end{equation}

\Cref{main_result} represents an important qualitative result, as it shows that the electrons can retain a significant portion of the energy they gain from $E_\parallel$. We find the terminal momentum gain by taking the limit of $x/x_R \rightarrow \infty$ in \cref{main_result}. There is a profound difference between odd and even radial radial modes, i.e. odd and even radial indices $p$. In the case of even modes with (e.g. $p = 0, 2, 4$), we have 
\begin{equation} \label{eq:p_term}
    \Delta p^{term}_\parallel = \frac{|e| E_{\parallel}^{\max} x_R}{(p+1)c}  \cos (\Phi_0 -\pi).
\end{equation}
The energy gain occurs for \rc{$0.5\pi <\Phi_0 < 1.5\pi$} regardless of the radial mode structure. Our assumption that the electron is moving forward with an ultra-relativistic velocity breaks down for \rc{$1.5\pi/ <\Phi_0 < 2.5\pi$}, which invalidates \cref{eq:p_term} for these injection phases. In contrast to the even radial modes, there is no terminal momentum gain, $\Delta p^{term}_\parallel = 0$, for the odd radial modes regardless of the injection phase. It is worth pointing out that this estimate was obtained under the assumption that the electron is ultra-relativistic, so the analysis has to be revised along the parts of the trajectory where $c - v_x \geq v_{ph} - c$. However, the value of $\Delta p^{term}_\parallel$ is unlikely to increase dramatically as a result.

\begin{figure}
 \centering
 \includegraphics[width=0.59\linewidth]{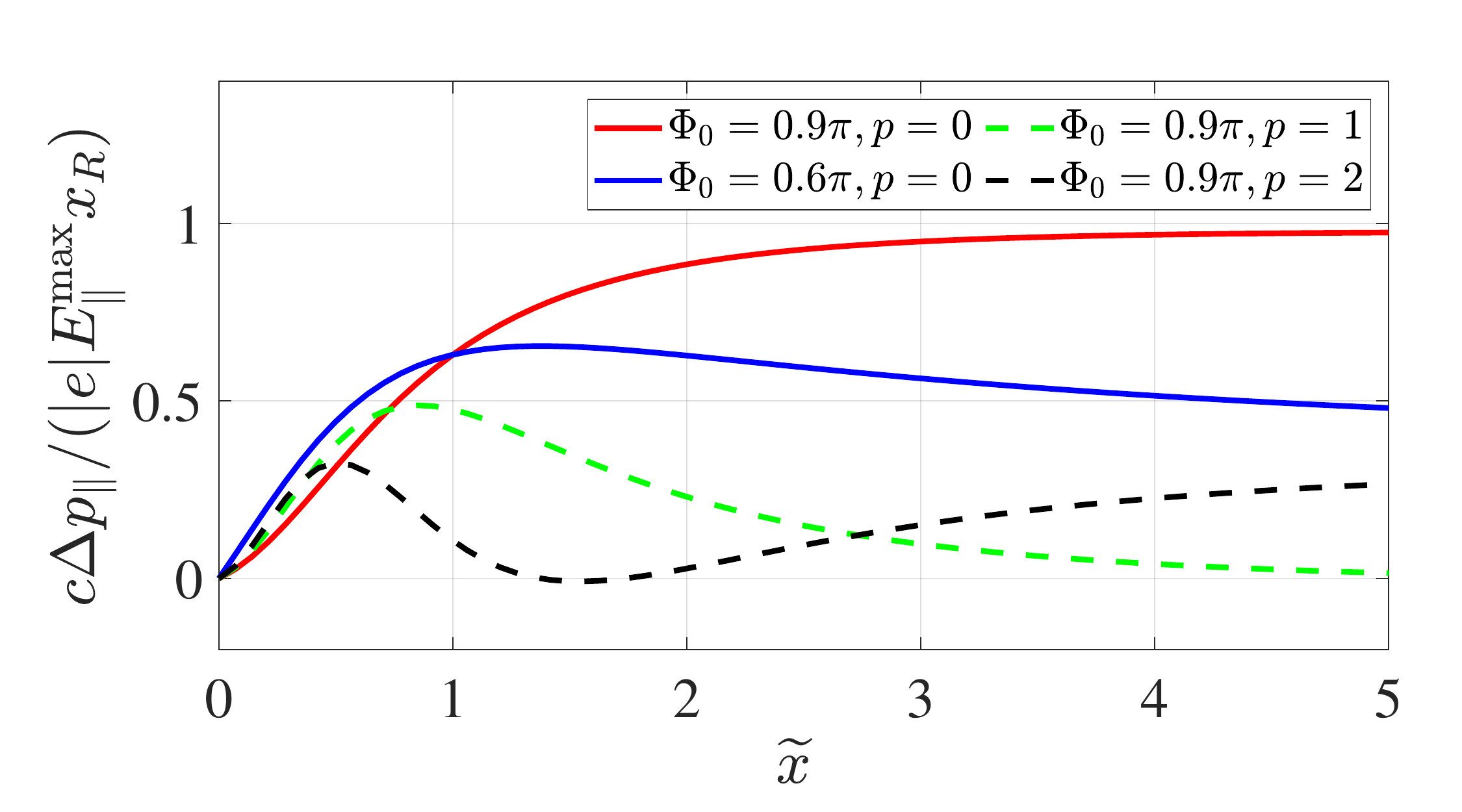} %LG-CP Detailed paper/Delp_acc.pdf
\caption{The momentum gain predicted by \cref{main_result} for electrons injected at phase $\Phi_0$ into a right-circularly polarized beam ($\sigma = 1, l = -1$).  The solid curves are for a beam with the radial index $p=0$, whereas the green and black dashed curves correspond to beams with $p=1$ and $p=2$, respectively.} \label{ExAcc_discuss}
\end{figure}

\Cref{ExAcc_discuss} illustrates the results of our analysis for three different radial modes, $p = 0, 1,$ and 2. The solid curves show electron acceleration by a beam with $p=0$, as predicted by \cref{main_result}. Both electrons are injected at a phase that leads to a net momentum gain. However, the delayed injection of the electron shown with the solid blue curve \rc{($\Phi_0 = 0.6\pi$)} means that it slips into the decelerating phase before the amplitude of $E_\parallel$ becomes negligibly small due to the beam diffraction. As a result, the net momentum gain is two times lower than for the case of \rc{$\Phi_0 = 0.9\pi$}. Higher-order modes speed up the electron slip into the decelerating phase, because the relative degree of superluminosity given by \cref{eq:superlum} increases with $p$. This trend is clearly shown by \rc{the dashed curves}, representing modes with $p=1$ and $p=2$, respectively. In both cases, the injection phase is \rc{$\Phi_0 = 0.9\pi$}, which is the same phase as that for the solid red curve ($p=0$).  \rc{The green dashed curve}, corresponding to $p=1$, rolls over{, after a peak in momentum gain,} at $\widetilde{x} \approx 1$, whereas \rc {the black dashed curve}, corresponding to $p=2$, rolls over even sooner when the electron reaches $\widetilde{x} \approx 0.5$. In the case of $p=1$, the electron remains in the decelerating phase until the beam experiences significant diffraction. As a result, there is no net momentum gain. In the case of $p=2$, the dephasing is faster, so the electron enters another accelerating region of the beam when it reaches $\widetilde{x} \approx 1.5$. The acceleration continues until the diffraction eliminates $E_\parallel$. In this case there is a net momentum gain, but it is smaller than in the case of $p=0$ because the acceleration time is shorter and the accelerating field (due to the diffraction) is weaker.

It is useful to recast our results for the momentum gain in terms of electron energy. The energy of an ultra-relativistic electrons with momentum $\bm{p}$ is $\varepsilon \approx c |\bm{p}|$ and, in our case, $|\bm{p}| \approx p_\parallel$. If the electron experiences a significant momentum gain, then the terminal longitudinal electron momentum is $p_\parallel^{term} \approx \Delta p_\parallel^{term}$. Putting all these estimates together, we find that the terminal electron energy is given by $\varepsilon^{term} \approx c \Delta p_\parallel^{term}$. For modes with an even $p$ index, the momentum gain is given by \cref{eq:p_term}. We also take into account the relation between $a_{\parallel}$ and $P$ given by \cref{a and P}. As a result we arrive 
to the following expression for the electron energy gain:
\begin{equation} \label{max energy}
 \varepsilon^{term} [\mbox{GeV}] \approx \frac{0.72}{p+1} \cos (\Phi_0 - \pi) P^{1/2} [\mbox{PW}], 
\end{equation}
The result is independent of the spot size $w_0$ and wavelength $\lambda_0$. At the end of \cref{sec:estimates}, we examined the field structure of a circularly-polarized 0.6~PW laser beam with $l = -1$, $\sigma = 1$, and $p = 0$. According to \cref{max energy}, we expect this beam to generate electrons with hundreds of MeV in energy, since \rc{ $\varepsilon^{term} \approx 0.56 \cos(\Phi_0 - \pi)$~GeV.} 

%++++++++++++++++++++++++++++++++++++++++++++++
%++++++++++++++++++++++++++++++++++++++++++++++

\section{Simulation results for a normally incident 600~TW laser}\label{Sec-sim}

\begin{figure}
 \centering
 \includegraphics[width=0.85\linewidth]{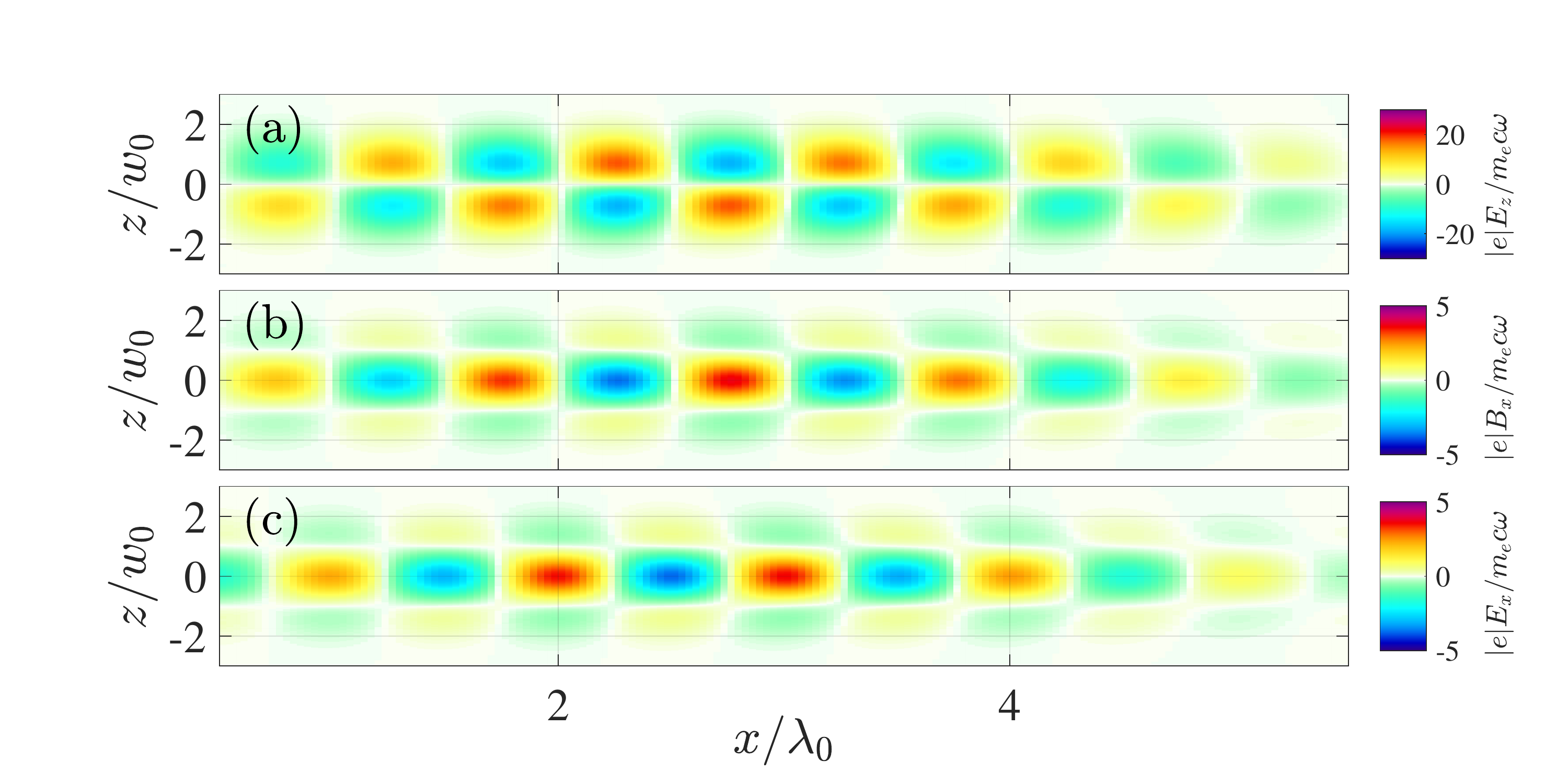}
\caption{Field structure of an incident circularly-polarized Laguerre-Gaussian laser beam  from a 3D PIC simulation. The snapshots are taken at $t = -9$~fs. The laser beam and simulation parameters are given in \cref{table:PIC}.
} \label{eb_xz}
\end{figure}
% The fields are normalized to $E_0 = B_0 \equiv m_e c \omega/|e|$.

In \cref{sec:estimates}, we provided preliminary estimates showing that a properly chosen beam with twisted \rc{wavefronts} can generate forward-directed ultra-relativistic electrons. In this section, we present a self-consistent analysis, performed using a 3D PIC simulation, of electron injection and subsequent acceleration by a laser beam with dominant $E_\parallel$ and $B_\parallel$ in the near-axis region.

\begin{table}
\centering
\begin{tabular}{ |m{7cm}|m{5cm}| }
 \hline
 \multicolumn{2}{|l|}{\textbf{Laser parameters} }\\
 \hline
 Peak power (period averaged) & 0.6~PW \\
 Twist and radial index & $l = -1, p = 0$ \\
 Right-circular polarization & $\sigma = 1$ \\
 Wavelength & $\lambda_0 = 0.8~\micron$\\
 Pulse duration ($\sin^2$ electric field) & $\tau_g=20$~fs\\
 Focal spot size ($1/\mathrm{e}$ electric field) & $w_0 = 3~\micron$\\
 Location of the focal plane & $x = 0~\micron$\\
 Direction of the incident laser & $-x$ \\
 \hline \hline
 \multicolumn{2}{|l|}{\textbf{Other parameters} }\\
 \hline
 Position of the bulk target & $-1.0~\micron \leq x \leq -0.3~\micron$\\
 Position of the pre-plasma & $-0.3~\micron < x \leq 0.0~\micron$\\
 Electron and C$^{+6}$ density & $n_e = 500\;n_{c}$ and $n_{Carbon} = 83.3\;n_{c}$ \\
 Gradient length & $L = \lambda_0$/20\\
 Simulation box ($x \times y \times z$) & $10~\micron \times 30~\micron \times 30~\micron$\\ 
 Moving window start time & 11 fs \\
 Moving window velocity & $ c$ \\
 Cell number ($x \times y \times z$) & 400 cells $\times$ 800 cells $\times$ 800 cells \\
 Macroparticles per cell for electrons & 300 at $r < 2.5~\micron$, 36 at $r > 2.5 \micron$ \\
 Macroparticles per cell for C$^{+6}$ & 24 \\
 %Macroparticles per cell for C$^6$+ & 28 \\
 \hline 
 \end{tabular}
 \caption{3D PIC simulation parameters.}
 \label{table:PIC}
\end{table}

In our simulation, a 600~TW circularly-polarized Laguerre-Gaussian beam with $l = -1$, $\sigma = 1$, and $p = 0$ is normally incident on a mirror that is initialized as a fully ionized plasma with a sharp density gradient. The incident pulse propagates in the negative direction along the $x$-axis. \rc{The laser envelope function $g({\xi})$ has a temporal profile such that $g(t)=\sin^2(0.5\pi*t/\tau_g)$ with a total duration of $\tau_g = 20$~fs.}  The beam width is $w_0 = 3~\micron$, the laser wavelength is $\lambda_0 = 0.8~\micron$, and the focal plane is located at $x = 0~\micron$. The mirror is a carbon plasma with an electron density profile $n_e = 500 n_c \exp[ - 20 (x + 0.3~\micron)/ \lambda_0]$ at $x \geq -0.3~\micron$, where $n_c = 1.8 \times 10^{27}$~m$^{-3}$ is the critical density for $\lambda_0 = 0.8~\micron$. While ion mobility does not appear to affect the simulation results, the ions are left mobile so as to ensure a more realistic scenario. The initial kinetic energy of all particles (electrons and ions) is set to zero. {In order to follow the electrons bunches over a long period of time a moving window is employed. The window size is set to encompass the entire simulation box and moves at a velocity $ c$, beginning at a time $t=$11 fs .} Additional simulation details are provided in \cref{table:PIC}. 

The field structure of the incident laser beam in the $(x,z)$-plane is shown in \cref{eb_xz}. The time snapshots are taken at $t = - 9$~fs. We define $t = 0$ as the time when, in the absence of the mirror, the peak of the laser envelope reaches at $x = 0$. As can be seen in \cref{eb_xz}, the longitudinal fields $E_x$ and $B_x$ are strongest on-axis where the transverse field $E_z$ tends to zero. The field structure in these snapshots agrees with the paraxial analysis presented in \cref{Sec-2} for beams with twisted \rc{wavefronts}. %%It is important to note that only the transverse fields are specified in the PIC input parameters and that any measured longitudinal fields are as a result of the Maxwell solver in the PIC code. 

\begin{figure}
 \centering
 \includegraphics[width=0.42\linewidth]{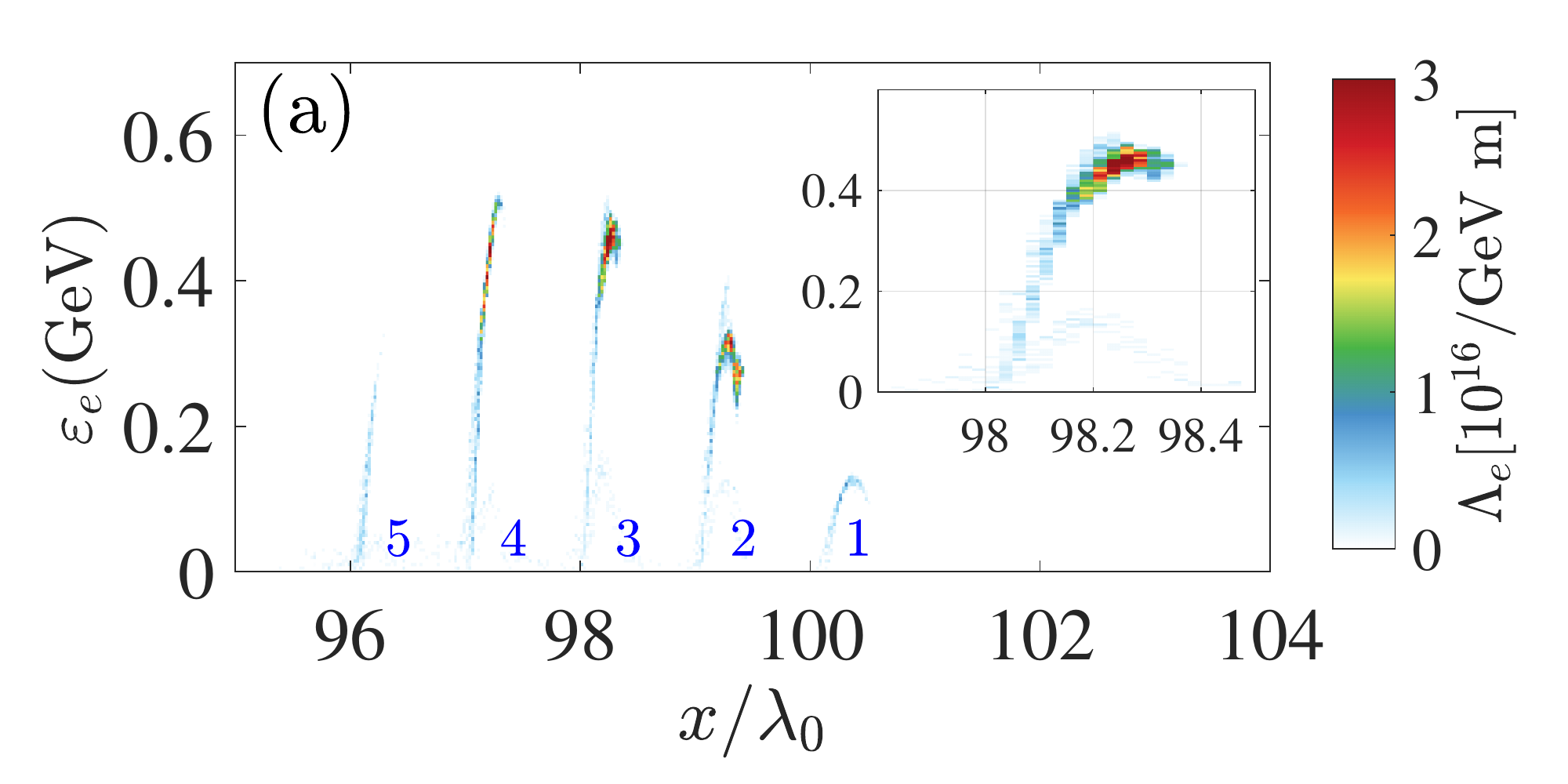} %flgcp5 PhaSp1_0300.pdf
 \includegraphics[width=0.42\linewidth]{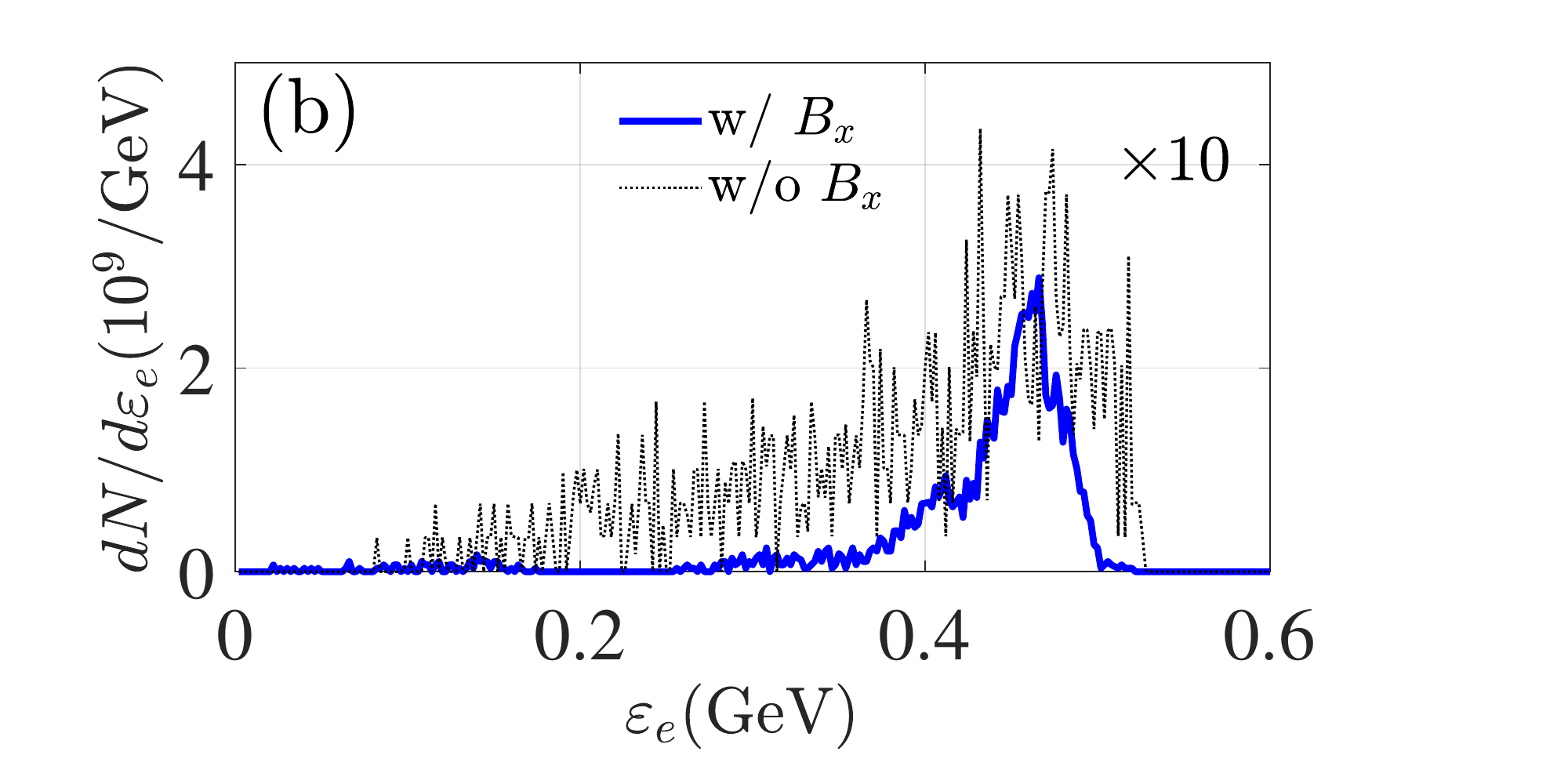} %LG-CP Detailed paper  Espec_nob0300.pdf
  \includegraphics[width=0.42\linewidth]{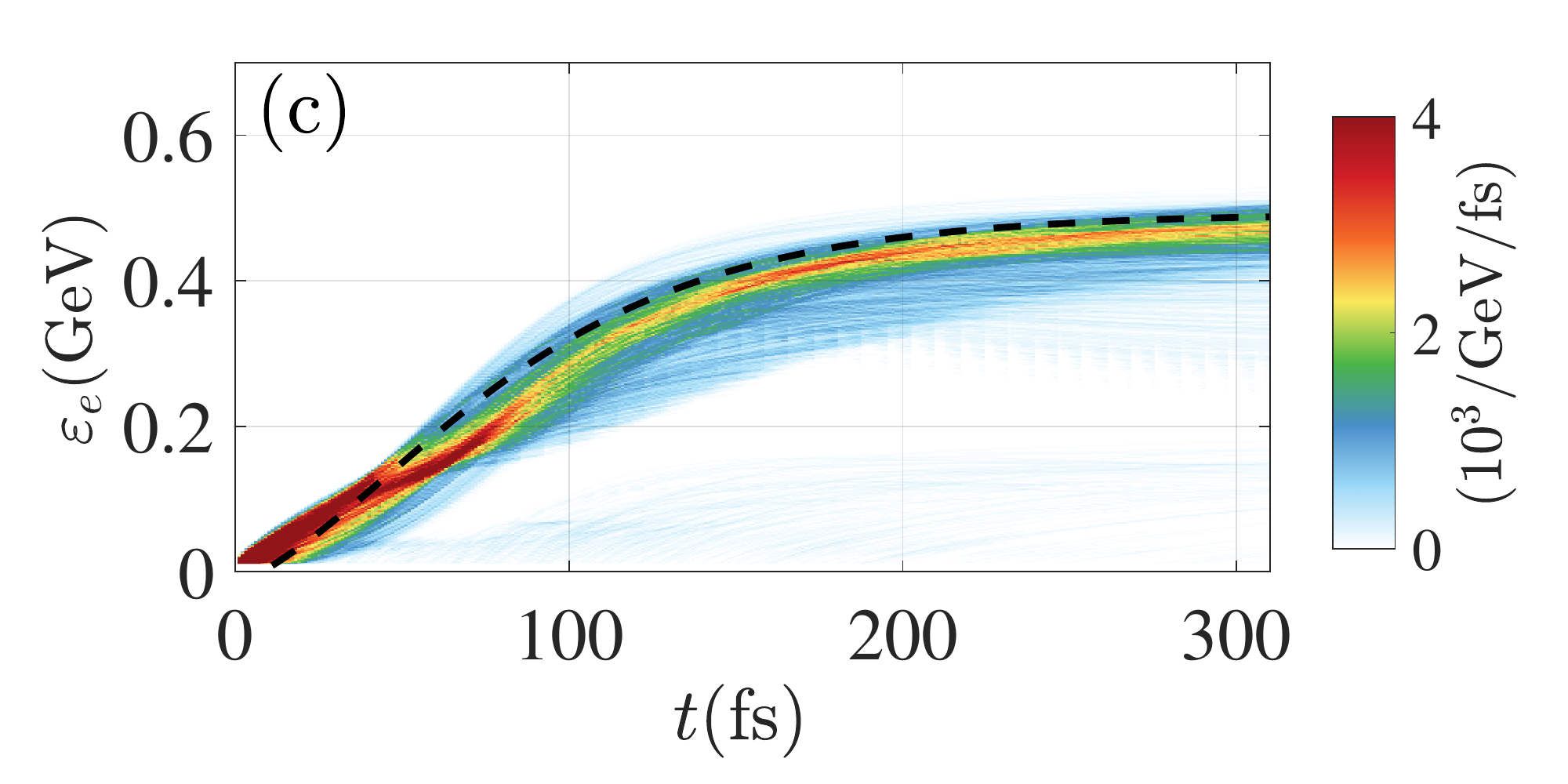} %flgcp5  Eng_Time3_0349.pdf
 \includegraphics[width=0.42\linewidth]{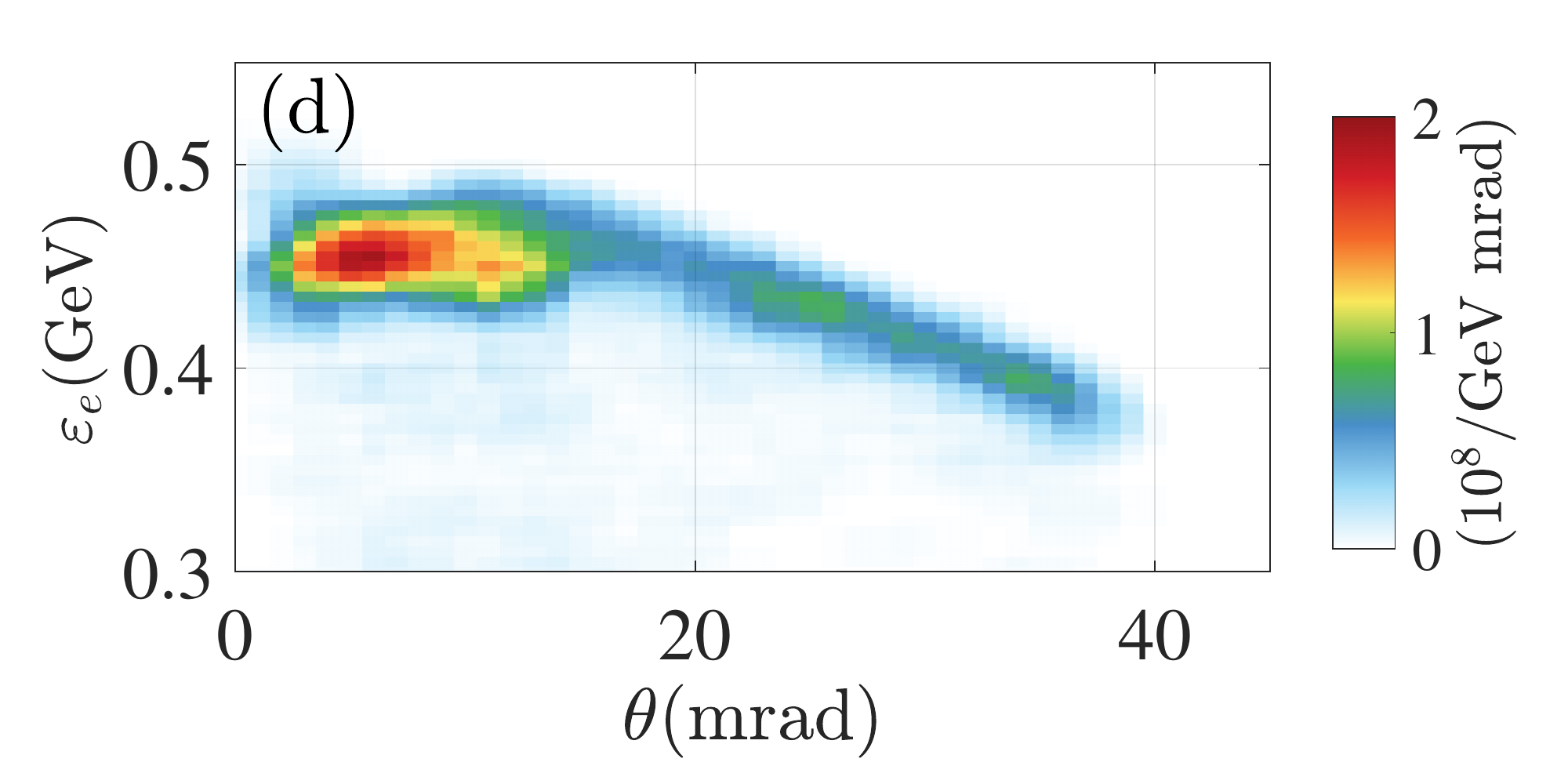} %flgcp5   EngAng2d1_hp_0300.pdf
\caption{(a) Accelerated electron bunches at $t = 261$~fs. The color shows the distribution of electrons with $r < 1.5~\micron$ in the $(\varepsilon_e, x)$-space. The inset shows the structure of the third bunch. (b)~The energy spectrum $dN/d\varepsilon_{e}$ in the third bunch at $t = 261$~fs for electrons with $r < 1.5~\micron$ for cases with (blue line) or without (black line, multiplied by 10 to aid the comparison) the force from $B_x$ in the electron equation of motion. (c)~The time evolution of the energy spectrum of the third bunch. The black dashed curve is the prediction of \cref{main_result} with  \rc{$\Phi_0 = 0.8\pi$}. (d)~The energy-angle distribution in the third bunch at $t = 261$~fs.} \label{ebunch}
\end{figure}

{The laser-plasma interaction that occurs} after the laser reaches the mirror can be sorted into two stages: the injection stage and the acceleration stage. The injection stage occurs during laser reflection off the plasma mirror. At this stage, plasma electrons are drawn out of the mirror by laser fields and injected into the near-axis region of the laser beam in front of the mirror. The electron injections stops after the laser beam is fully reflected off the mirror. \rc{Results from tracking the particles in the third bunch back to the target surface show a complex relationship between the transverse and longitudinal field structures. This process requires significant further research that is beyond the scope of this article.} 
%\rc{Our results show that the injection process is involving both the transverse field structure and the longitudinal magnetic field structure. This needs further study.}
During the acceleration stage, the injected electrons travel with the reflected laser beam away from the surface of the mirror. Their longitudinal motion is caused by $E_\parallel$ of the reflected laser beam. Only those electrons that are injected into a region with $E_\parallel < 0$ can accelerate while moving away from the mirror. On the other hand, no acceleration is possible when $E_\parallel > 0$ in the vicinity of the mirror surface. As a result, oscillations of $E_\parallel > 0$ near the mirror surface generate electron bunches rather than a continuous electron beam. The maximum areal density of a bunch, $\rho_e$, (integral of $n_e$ along the bunch) can be estimated by taking into account that the injection process during one laser period stops once the space-charge of the extracted electrons shields $E_\parallel$ of the laser. This yields $\rho_e \approx a_{\parallel} n_c c / \omega$ or 
\begin{equation}
    \rho_e [\mbox{m}^{-2}] \approx 1.3 \times 10^{22} P^{1/2}~[\mbox{PW}]~\lambda_0~[\mathrm{\mu m}]~w^{-2}~[\mathrm{\mu m}].
\end{equation}
The implication is that a beam with a strong longitudinal electric field is expected to produce high density bunches.

\Cref{ebunch}(a) shows that the accelerated electrons indeed travel in bunches. The snapshot is taken at $t = 261$~fs after the injected electrons have travelled roughly $100 \lambda_0$ along the $x$ axis with the laser beam. The plot shows the distribution of the electrons located in the region near the axis with $r < 1.5~\micron$ in the $(\varepsilon_e,x)$-space. The compactness of the bunches, both in transverse [see \cref{eDenst_Injct}(c)] and longitudinal directions [see \cref{ebunch}(a)], ensures that the electrons in each bunch essentially experience the same accelerating field. This explains the narrow energy spread within each bunch in \cref{ebunch}(a).

The evolution of the energy spectrum within a single bunch is plotted in \cref{ebunch}(c). The selected bunch is marked as bunch \#3 in \cref{ebunch}(a). The bunch \rc{retains a relatively narrow energy spread} during the acceleration process. The dashed curve shows the prediction given by \cref{main_result} for an injection phase of \rc{$\Phi_0 = 0.8\pi$}. The good agreement indicates that the accelerated electrons are likely injected into the same phase of the reflected beam. There is also a strong correlation between electrons which have a high energy and those with a low divergence angle. This is best seen in \cref{ebunch}(d) which shows the distribution of energy versus divergence angle $\theta=\arctan{\left(p_{\perp}/p_\parallel \right)}$ of the third bunch. 

The presented simulation demonstrates that a normally incident 600~TW beam can generate \rc{several dense bunches of} ultra-relativistic electrons. Specifically, the terminal energy of the electrons in the third bunch is 0.47~GeV with a FWHM of $\sim10\%$ [see \cref{ebunch}(b) for a snapshot at $t = 261$~fs]. The bunch has a duration of roughly 400~as and a total electron charge of 26~pC, while the normalized transverse emittance is $9.5~\micron$. 

%%% time shift is 39fs. here is 48fs - 39fs

%%% time shift is 39fs. here is 300fs - 39fs

%++++++++++++++++++++++++++++++++++++++++++++++
%++++++++++++++++++++++++++++++++++++++++++++++

\section{The impact of the longitudinal magnetic field  on electron dynamics}\label{Sec-bx}

The most distinctive feature of the considered laser beam with twisted \rc{wavefronts} is the strong longitudinal magnetic field $B_\parallel$ in the near-axis region. In what follows, we examine the profound impact of this field on electron dynamics.

\begin{figure}
 \centering
 \includegraphics[width=0.95\linewidth]{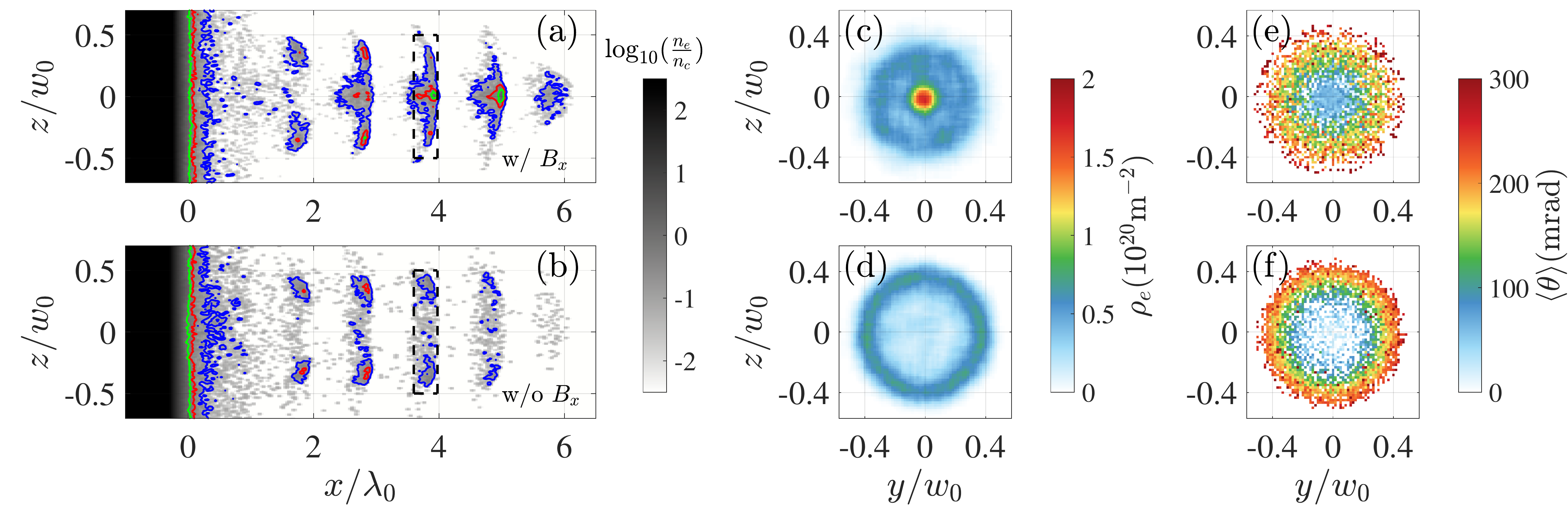} %LG-CP Detailed paper eDenst_ep3dw_2d0024.pdf
\caption{A comparison between 3D PIC simulations with (upper row) and without $B_x$ (bottom row) in the electron equations of motion. Both simulations use the same 600~TW laser beam with parameters given in \cref{table:PIC}. All snapshots are taken at $t = 9$~fs.
(a) and (b): Electron density on a log-scale, with the color representing $\log(n_e/n_c)$. The blue, red, and green contours denote $n_e = 0.1n_{c}$, $0.5 n_{c}$, and $n_{c}$. The dashed rectangle shows the radial and longitudinal extent of what is referred to as the third bunch in the remaining panels. (c) and (d): Areal density of the electrons in the third bunch. (e) and (f): Cell-averaged electron divergence angle $\langle \theta \rangle$ in the third bunch. The angle $\theta \equiv \arctan(p_{\bot}/p_{x})$ of an individual electron is averaged over the electrons located within the cells with the same $y$ and $z$ coordinates.} \label{eDenst_Injct}
\end{figure}

In addition to $E_\parallel$, the electrons in the near-axis region experience $B_\parallel$ that can provide transverse confinement at an early stage of the acceleration before the electron momentum becomes predominantly forward-directed. The importance of $B_\parallel$ can be assessed by estimating the Larmor radius, $r_L$, for an electron injected with transverse relativistic momentum $p_\perp$ into the field near the focal plane, whose strength is given by \cref{a and P}. We find that 
\begin{equation}
    r_L/w_0 \approx \frac{2.2 \times 10^{-3}}{\sqrt{1+p}} \left(\frac{p_\perp}{m_e c}\right) \left(\frac{w_0}{\lambda_0 }\right)  P^{-1/2}~[\mbox{PW}].
\end{equation}
In the case of the 600~TW laser beam considered in \cref{Sec-sim}, we have $r_L/w_0 \approx 10^{-2} p_\perp/m_e c$. Even for those electrons that are injected with $p_\perp/m_e c \approx |e| E_\perp^{\max} / m_e c \omega \approx 19$, the Larmor radius, $r_L \approx 0.2 w_0$, is significantly smaller than the beam waist. By keeping the injected electrons close to the axis, $B_\parallel$ ensures that the electrons are unable to sample strong $E_\perp$ and gain additional transverse momentum. As a result, they are predominantly accelerated by $E_\parallel$, which leads to a strongly collimated beam. It is important to point out that $B_\parallel$ is shifted by $\pi/2$ with respect to $E_{\parallel}$, which means that the direction of the electron rotation induced by the magnetic field changes during the acceleration process.

At the injection stage, plasma electrons experience a complicated interplay of transverse and longitudinal fields that is not captured by the presented estimate. The impact of $B_x$ during this process can be elucidated by removing its effects from the electron dynamics even though this does lead to a somewhat non-physical scenario. To remove the effect of $B_x$ in the PIC simulation, we multiply the $B_x$ variable by zero in the relevant section of the particle pusher. \Cref{eDenst_Injct}(b) shows the electron density obtained using this approach for the laser-mirror interaction examined in \cref{Sec-sim}. For comparison, \cref{eDenst_Injct}(a) also shows the electron density from the original simulation where the effect of $B_x$ is included. The bunches in \cref{eDenst_Injct}(b) form in a ring-like structure \textit{away} from the central axis. The structure of the bunches appears superficially similar to that observed for radially polarized beams at lower intensities~\cite{Zaim2017}. In contrast to that, the electron bunches in \cref{eDenst_Injct}(a) are formed close to the axis in the region where the longitudinal field dominates. The bunches become more compact as they move away from the mirror, with the electron density exceeding $n_c$.

\begin{figure}
 \centering
 \includegraphics[width=0.75\linewidth]{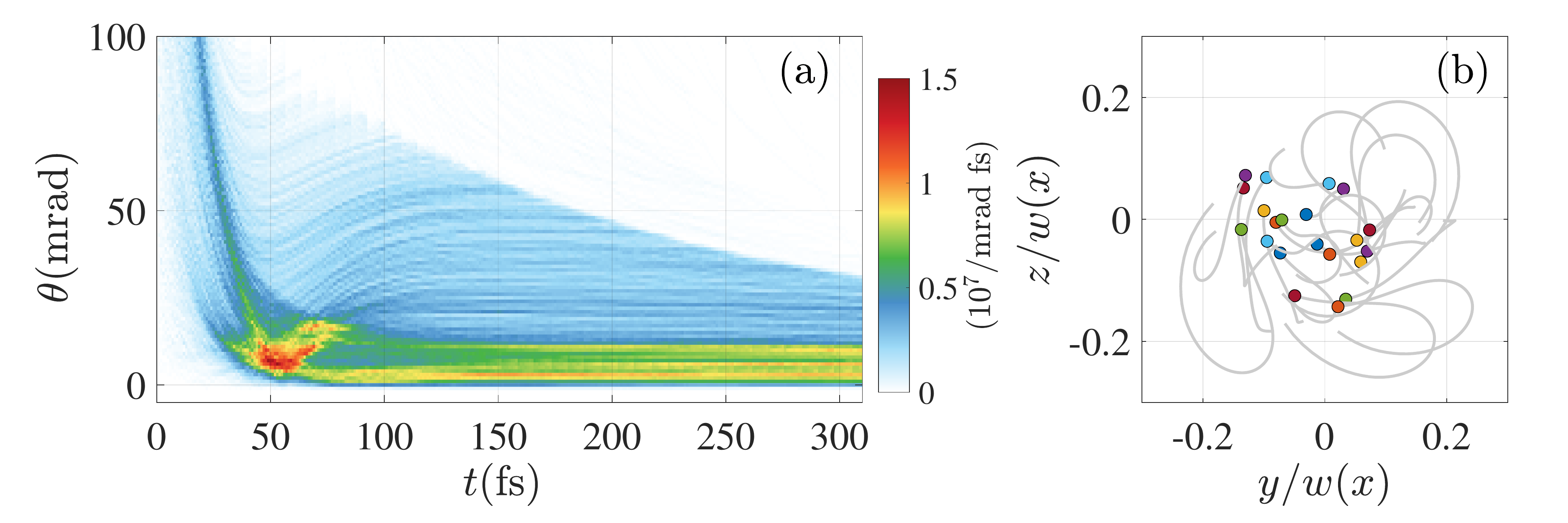} %E:\HaoHan\flgcp5  flgcp5_Ang_Time.pdf
\caption{(a) Time evolution of the electron distribution over the divergence angle $\theta$ in the third bunch ($r < 1.5~\micron$). (b) Trajectories of electrons randomly chosen from the third bunch at $t = 261$~fs. The trajectories correspond to $6 \mbox{ fs} \leq t \leq 261~\mbox{ fs}$, with the circles marking the electron positions at $t = 261$~fs. The transverse coordinates are normalized to the local width of the laser beam $w(x) = w_0 \sqrt{1 + x^2/x_R^2}$. } \label{eTraj_yz_wx}
\end{figure}

Figures~\ref{eDenst_Injct}(c)~-~(f) compare the areal density and divergence angle between the two simulations. The comparison is performed for the electrons within the third bunch in Figs.~\ref{eDenst_Injct}(a) and (b). The radial and longitudinal extend of the third bunch is shown with the the black dashed rectangles in Figs.~\ref{eDenst_Injct}(a) and (b). The difference in collimation is striking even at this injection stage ($t=9~\mathrm{fs}$). The higher areal density point-like structure seen in \cref{eDenst_Injct}(c) is complimented by a corresponding small cell-averaged electron divergence angle $\langle \theta \rangle$ seen in \cref{eDenst_Injct}(e). The averaging, indicated by the angular brackets, is performed by taking all the electrons located within the cells with the same $y$ and $z$ coordinates, i.e. the cells with the same projection onto the transverse plane. The averaging accounts for the difference in the numerical weight $w_\alpha$ of individual macro-particles caused by the initial density gradient: $\langle \theta \rangle = \sum_{\alpha} w_\alpha \theta_\alpha / \sum_{\alpha} w_\alpha$. In the case where the influence of $B_x$ is removed [with the corresponding plots of areal density in \cref{eDenst_Injct}(d), and cell-averaged divergence in \cref{eDenst_Injct}(f)], we see a ring-like `bunch' with divergent profile indicating that it will continue to diverge long after the injection stage is over. The contrasting pictures between the two simulations show the important role of the $B_x$.

\Cref{ebunch}(b) shows the energy spectrum of the third bunch in the simulation without the $B_x$ forces at a much later time ($t = 261~\mathrm{fs}$). The same figure shows the spectrum for the third bunch in the original simulation at the same time instant. Both spectra are calculated for the electrons in the near-axis region with $r < 1.5~\micron$. Not surprisingly, the electrons moving close to the central axis achieve similar energies in both simulations. However, the number of such energetic electrons in the simulation without the $B_x$ forces is an order of magnitude lower. During the injection process and at the very early acceleration stage the longitudinal magnetic field provides a significant amount of transverse electron confinement. This keeps electrons in the region with a strong $E_x$ but weak $E_{\perp}$, which ensures good collimation of the electron bunches.

To further examine the transverse electron motion during the acceleration process, we have tracked several electrons. The electrons are randomly selected from the third bunch ($r < 1.5~\micron$) at $t = 261$~fs. The projection of the electron trajectories onto the transverse plane for $6 \mbox{ fs} \leq t \leq 261~\mbox{fs}$ is shown in \cref{eTraj_yz_wx}(b), where the circles indicate the electron positions at $t = 261$~fs. The transverse coordinates are intentionally normalized to the local width of the laser beam $w(x) = w_0 \sqrt{1 + x^2/x_R^2}$ to correlate the electron location with the strength of the transverse electric field $E_\perp$ that increases away from the central axis. As expected, there is a pronounced rotation induced by $B_\parallel$ that prevents electrons from expanding and reaching a region with a strong $E_{\perp}$. This rotation is also observed to reverse, effectively twisting the trajectory of the electrons. This feature is caused by the $\pi/2$ shift between $E_\parallel$ and $B_\parallel$. The sign of $B_\parallel$ changes during the acceleration process, which causes the electrons to rotate in the opposite direction and manifests itself as the twist seen in \cref{eTraj_yz_wx}(b) for each individual trajectory.

Another important metric of the electron dynamics is the divergence angle $\theta$. \Cref{eTraj_yz_wx}(a) shows how the electron distribution over $\theta = \arctan{\left(p_{\perp}/p_\parallel \right)}$ changes with time in the third bunch.  We find that the bunch becomes less divergent over time, which is a clear indicator of the dominant role played by $E_\parallel$. Indeed, $\theta$ decreases because $p_\parallel$ increases at a much faster rate than $p_\perp$. The increase in $p_\parallel$ is caused by $E_\parallel$, as already discussed in \cref{sec:estimates}. The changes in $p_\perp$ are caused by $E_\perp$. The longitudinal magnetic field has no direct impact on the divergence angle since it rotates $\bm{p}_\perp$ without changing its amplitude. However, by keeping the electrons close to the central axis, the magnetic field prevents them from experiencing a strong $E_\perp$ that peaks off axis and, as a result, it prevents $p_\perp$ from increasing.

\begin{figure}
 \centering
 \includegraphics[width=0.35\linewidth]{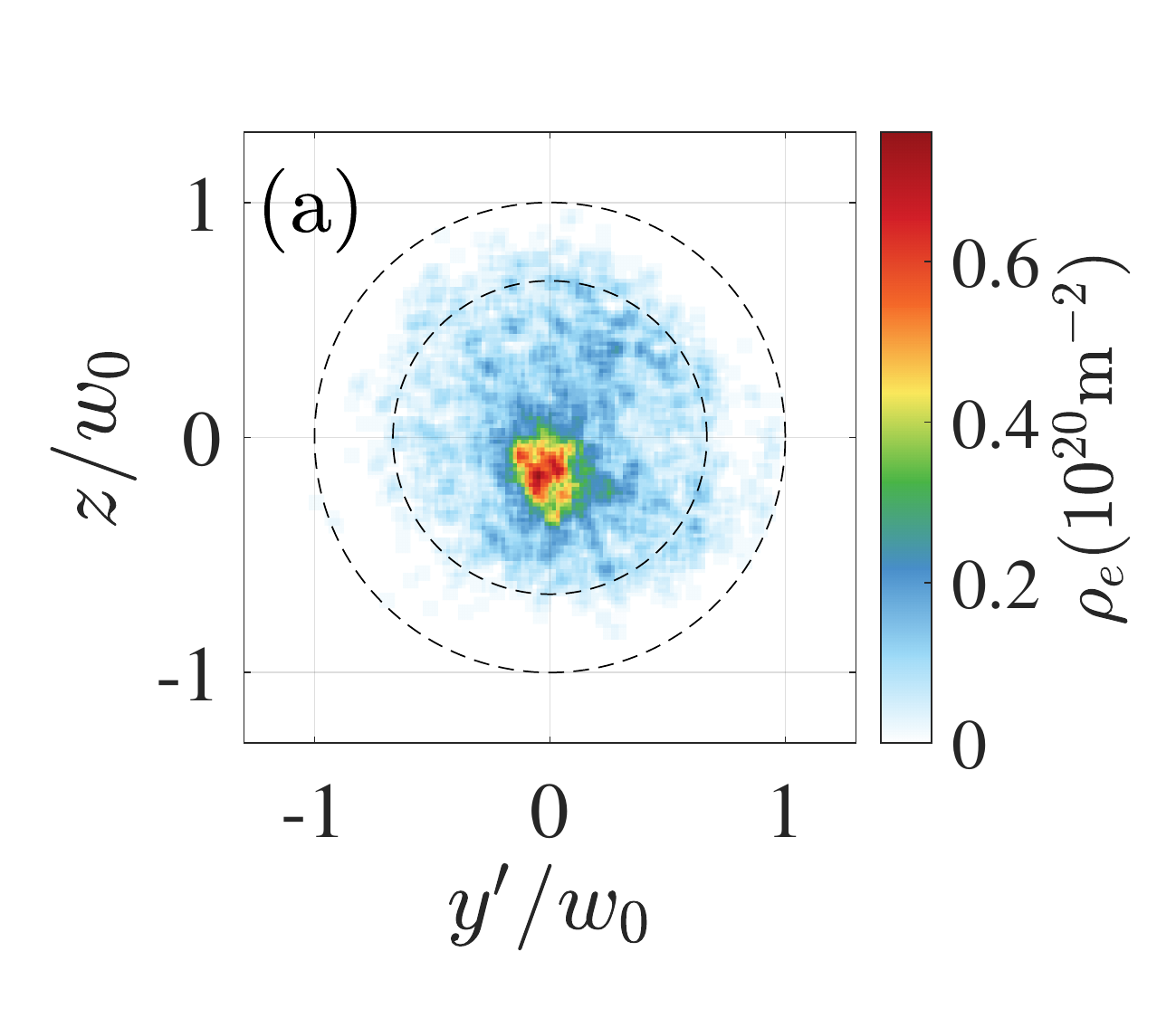} %flgcp5c_ep3dw1_2d_a_25ob0100.pdf
 \includegraphics[width=0.35\linewidth]{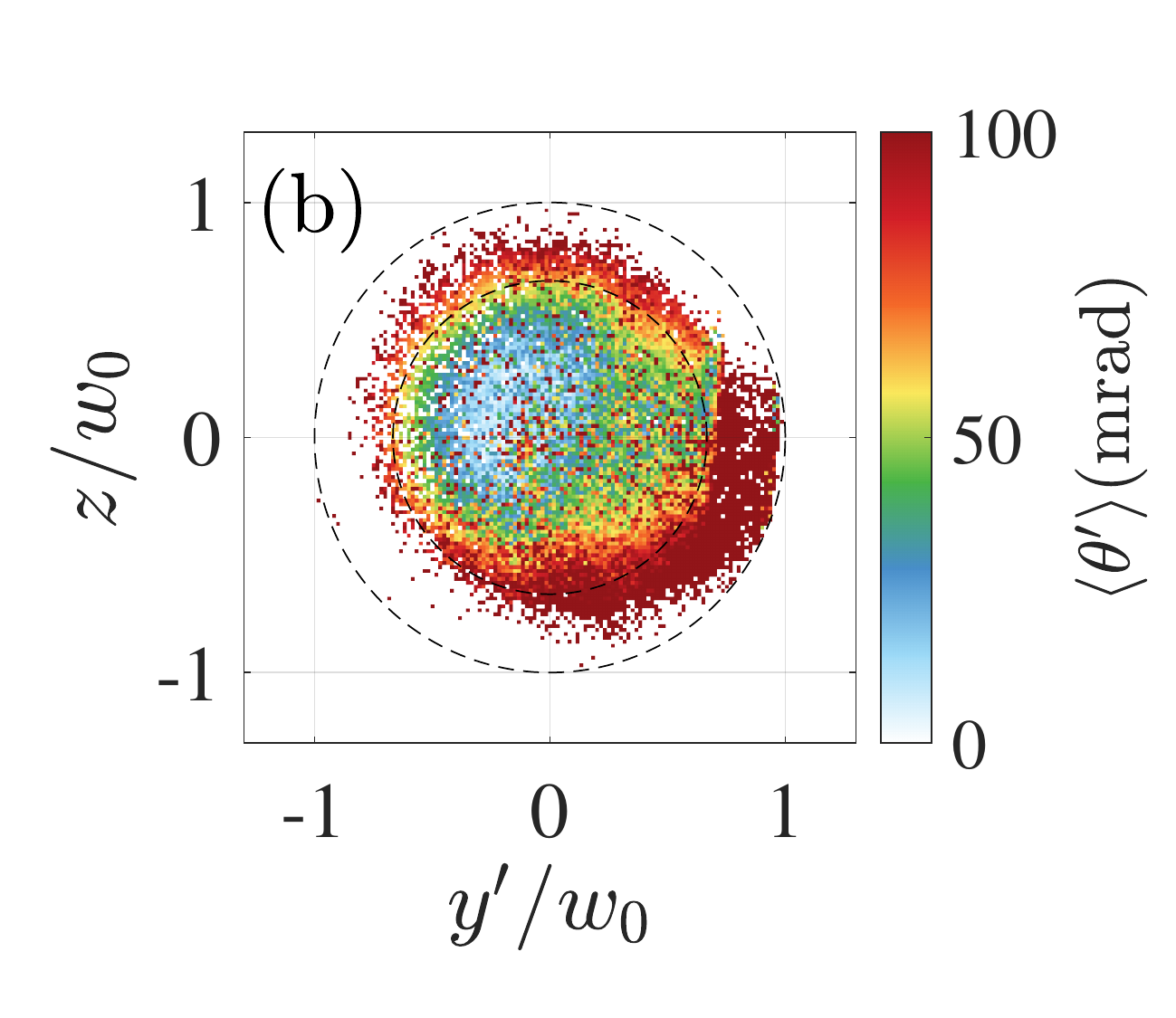} %flgcp5c_ep3dw1_2d_b_25ob0100.pdf
\caption{3D PIC simulation results for a $25^\circ$ incidence angle. The snapshots are take at $t = 39$~fs and plotted in an $(x',y',z)$ system of coordinates whose $x'$-axis points in the specular direction. (a)~Areal electron density in a third bunch moving in the specular direction. (b)~Cell-averaged electron divergence angle {$\langle \theta' \rangle$ in the third bunch. The angle $\theta' \equiv \arctan \left( \left. \sqrt{p_{y'}^2 + p_z^2} \right/ p_{x'} \right)$ is averaged on every mesh cell of the $(y',z)$ plane.}} \label{ebunch_2d_25ob}
\end{figure}

%\section{Simulation results with a oblique incidence} \label{Sec-5}

We conclude this section by presenting results for a case of oblique incidence. Experimental facilities often require that laser pulses are not shot at normal incidence onto a reflecting surface so as to avoid damage to optical systems. This was our primary motivation to examine the oblique incidence case. In our simulation, the laser beam is still incident in the negative direction along the $x$-axis, but the target is now titled by $25^\circ$. In order to make the simulation manageable, we perform it in a window moving along the $x$-axis. The simulation time is limited compared to the case of normal incidence, because the electron bunches, that now travel at an angle to the $x$-axis, leave the simulation window prematurely. \rc{This setup does not allow enough time for the acceleration process to reach a plateau as in the normal incidence case and so the final distribution function is left for future research. Despite this limitation this simulation will allow testing the early time conditions that set up the acceleration process and bunch formation.}

We find that the laser reflection again generates a series of dense collimated bunches, which indicates the robustness of the considered approach to electron acceleration. The areal density $\rho_e$ and the cell-averaged divergence angle $\langle \theta' \rangle$ in the third bunch are shown in \cref{ebunch_2d_25ob}(a, b). These snapshots are taken at \rc{$t =  52~\mathrm{fs}$}. We use an orthogonal $(x',y',z)$ system of coordinates whose $x'$-axis points in the specular direction. The divergence angle $\theta' \equiv \arctan \left( \left. \sqrt{p_{y'}^2 + p_z^2} \right/ p_{x'} \right)$ is the angle between the momentum vector $\bm{p}$ and the $x'$-axis. Even though the axial symmetry is broken, there is still strong evidence of a dense non-divergent bunch close to the laser axis. The high degree of collimation following the injection suggests that the reflected laser pulse would generate highly energetic bunches with similar characteristics to the case presented at normal incidence.

%++++++++++++++++++++++++++++++++++++++++++++++
%++++++++++++++++++++++++++++++++++++++++++++++

\section{Summary and discussion} \label{Sec-6}

In summary, this manuscript presents a detailed analysis of the topology of linearly and circularly polarized Laguerre-Gaussian laser beams. It is shown that the beams with a twist index $|l| = 1$ have a distinct field structure in the near-axis region with dominant $E_\parallel$ and $B_\parallel$. In the case of circularly polarized beams, the rotation of $\bm{E}_\perp$ should be in the opposite direction to the rotation of the \rc{wavefronts}, i.e. $\sigma = -l$, to achieve such a structure. The manuscript also presents kinetic 3D PIC simulations for a 600~TW circularly polarized beam ($p = 0$, $l = -1$, and $\sigma = 1$) reflected off a plasma mirror. The dominant $E_\parallel$ and $B_\parallel$ combine to inject dense electron bunches upon reflection. The bunches are effectively accelerated by $E_\parallel$ while being confined by $B_\parallel$. The magnetic field prevents the bunch electrons from travelling too far radially outwards, so they sample a relatively weak transverse electric field and remain well-collimated. \rc{The bunch with the largest energy has a distinctly narrow energy spread with a FWHM of just 10\%}. 
%\rc{One bunch is narrow energy spread due to their compactness, with energy in the range of 0.5~GeV (10\% FWHM energy spread)}. 
The terminal energy gain for an individual bunch is well predicted by the analytical model developed in the manuscript. The charge of a single bunch in the simulation is as high as 26~pC. The bunches have a duration of $\sim 400$~as and a remarkably low divergence of just \dg{1.15} (20~mrad).

The nearest analogue to the discussed mechanism is the acceleration by a radially polarized laser beam. Such a beam also has a strong longitudinal electric field that dominates the field structure in the region close to the central axis. However, in contrast to the beams considered in this manuscript, the radially polarized beam lacks a strong longitudinal magnetic field in the region close to the axis. As shown in \cref{append:RpDriven}, the absence of $B_\parallel$ has a profound impact on the electron acceleration even though the amplitude of $E_\parallel$ is the same as in the case of the radially polarized beam considered in \cref{Sec-sim}. In the absence of $B_\parallel$, there is no mechanism confining electrons in the region with $r/w(x) \ll 1$. The lack of confinement also manifests itself in the case of oblique incidence. \rc{We have shown some results that indicate the robustness of the electron injection by a beam with a strong $B_\parallel$.} In contrast to that, the injection by radially polarized beams is not as robust~\cite{Zaim2020}, which makes experimental implementation extremely challenging. 

{The transverse fields do have a role where they are, in-part, responsible for detaching electrons from the pre-plasma. The exact dynamical process relating to the transverse fields is likely of less importance than that of the longitudinal fields. Previous studies have discussed the ponderomotive force as a mechanism for keeping electrons~\cite{Xu2021} (and ions \cite{Pae_2020}) within the central region of the beam. However, as can be seen in the comparison made in \cref{eDenst_Injct}, this does not appear to be the dominant mechanism confining the electrons close to the axis in our case, at least at early times.}

\rc{While this study is limited to a relatively short laser pulse with 3 peaks at high amplitude, a TiSaphire laser with a FWHM of around 30 fs will have a train of roughly 15 peaks. The energy of each bunch is strongly tied to the amplitude at the position in the envelope. After being injected, electrons are accelerated directly by laser field in vacuum, implying that they are not likely subjected to instabilities that occur in bulk plasmas. The density of the bunches is also related to the longitudinal field, with higher densities recorded with higher amplitudes\cite{Shi2021}}

Using the scheme detailed in the manuscript, it may be possible to construct a source of highly collimated dense attosecond bunches of ultra-relativistic electrons suitable for potential applications. This can be achieved using a combination of high-power laser systems, similar to those already in use today. In addition to this, optical techniques that have already been put to experimental use can be used to create twisted wave-fronts necessary for this mechanism. Given the research presented here, with some further studies, a working design for a small-scale high-energy electron accelerator may be within grasping distance.

%++++++++++++++++++++++++++++++++++++++++++++++
%++++++++++++++++++++++++++++++++++++++++++++++

\section*{Acknowledgements}

Y.S. acknowledges the support by grant of KY2140000018 and Newton International Fellows Alumni follow-on funding. {D.R.B. and A.A. acknowledge the support by the National Science Foundation (Grant No. PHY 1903098).} Simulations were performed with EPOCH (developed under UK EPSRC Grants EP/G054950/1, EP/G056803/1, EP/G055165/1 and EP/ M022463/1). The simulations and numerical calculations in this paper have been done on the supercomputing system in the Supercomputing Center of University of Science and Technology of China. This work also used the Extreme Science and Engineering Discovery Environment (XSEDE), which is supported by National Science Foundation grant number ACI-1548562.

%++++++++++++++++++++++++++++++++++++++++++++++
%++++++++++++++++++++++++++++++++++++++++++++++

\appendix
\section{\sffamily{3D PIC simulation results for electron acceleration by a radially polarized laser beam}} \label{append:RpDriven}

The acceleration scheme presented in this manuscript is superficially similar to that associated with radially polarized beams~\cite{Zaim2017, Zaim2020}. Near the axis, radially polarized beams also have a strong longitudinal electric field, so a comparison of the two schemes is worth discussing.

\begin{figure}
 \centering
 \includegraphics[width=0.85\linewidth]{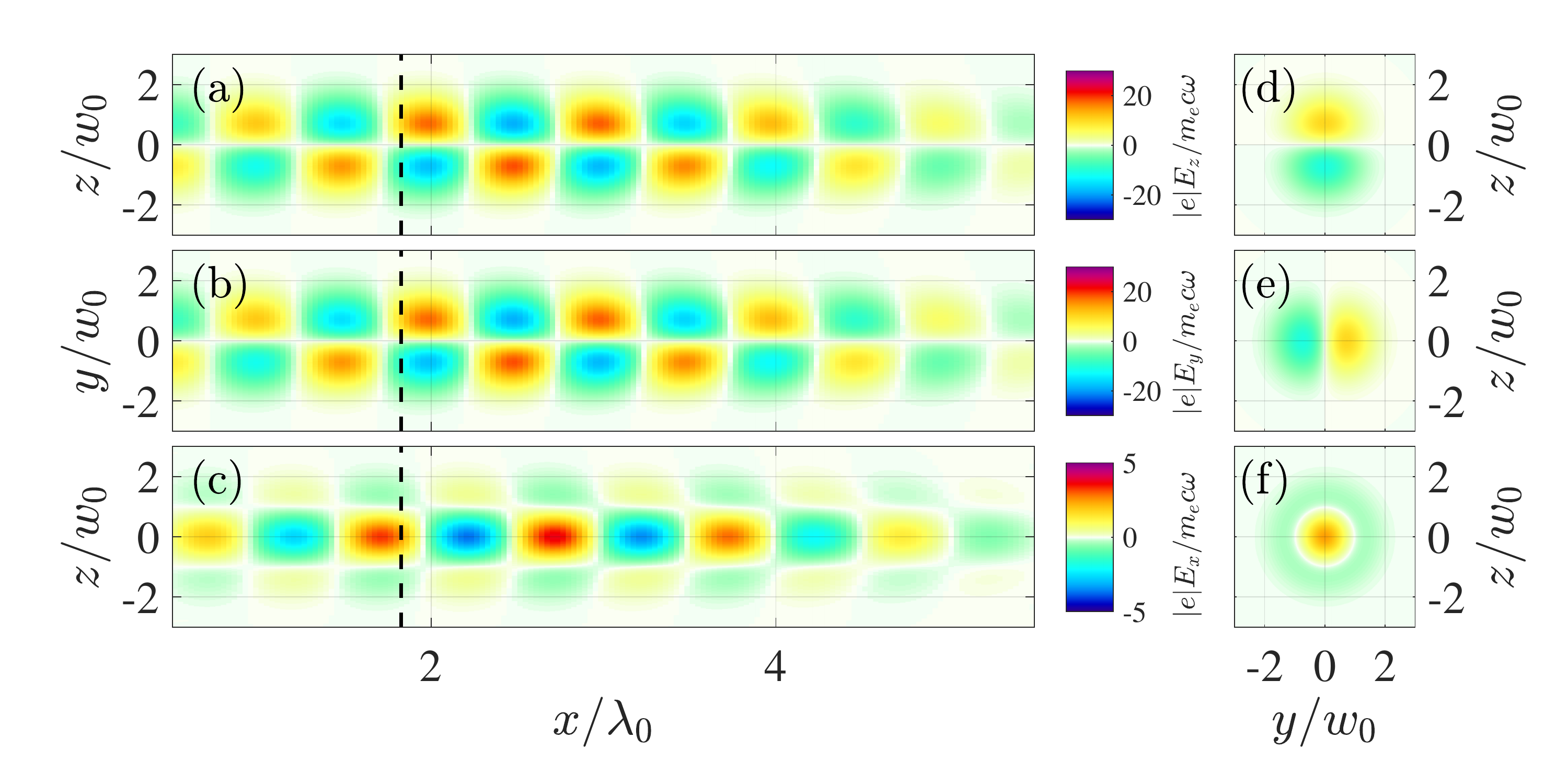} %frp2_Eb_one_fig0016.pdf   TACC-cn\flgcp5rp2
\caption{Electric field components of a radially polarized laser beam \emph{before} it encounters the mirror . Panels (a) and (d) show $E_z$; panels (b) and (e) show $E_y$; panels (c) and (f) show $E_x$. The left column [(a), (b), and (c)] shows the field structure in the $(x,z)$-plane at $y = 0$. The right column [(d), (e), and (f)] shows the field structure in the $(y,z)$-plane at the $x$-position indicated with the dashed line. All the snapshots are taken a time $t \approx -9$~fs. } \label{rp_eb}
\end{figure}

The difference in the field topology can be illustrated by constructing a radially polarized beam using two circularly-polarized Laguerre-Gaussian beams from \cref{Sec-2}: the first with $l = 1$, $\sigma = -1$ and the second, a mirror image of the first, with $l = -1$, $\sigma = 1$. We set $E_0 = E_*/2$ for each of the beams in \cref{E_y_1} that describes the structure of $E_y$. The resulting beam structure will be compared to a circularly polarized beam with $E_0 = E_*$ and $l = -1$, $\sigma = 1$, and $p=0$. The superposition of the two beams has
\begin{equation}
    E_y = \frac{1}{2} E_* \left( e^{i \phi} + e^{-i \phi} \right) D = E_* \cos \phi D,
\end{equation}
where $D(\widetilde{x},\widetilde{r},\xi) = g(\xi) \exp(i \xi) \psi_{\pm 1,p} (\widetilde{x},\widetilde{r},\phi) \exp(\mp i\phi)$ is a function that, for compactness, incorporates all the remaining dependencies besides the dependence on $\phi$. We next take into account that $E_z = i \sigma E_y$ in each of the circularly-polarized beams and find that the superposition of the two beams has
\begin{equation}
    E_z = \frac{1}{2} E_* \left( -i e^{i \phi} + i e^{-i \phi} \right) D = E_* \sin \phi D.
\end{equation}
These $E_z$ and $E_y$ represent a radially polarized laser beam:
\begin{equation}
    \bm{E}_\perp = \bm{e}_r E_* D.
\end{equation}

In the context of the electron acceleration mechanism, the biggest difference between the radially polarized beam and the circularly polarized beam with twisted \rc{wavefronts} is the absence of $B_\parallel$ near the axis. In order to find the structure of the longitudinal fields, we use Eqs.~(\ref{E_z_pm1}) and (\ref{B_z_pm1}). Close to the axis, the longitudinal electric fields of the two beams are the same, so the longitudinal field of the radially polarized beam is given by
\begin{equation}\label{Rp-Ex}
    E_x = \frac{i \theta_d}{2} \frac{1}{\widetilde{r}} e^{-i \phi} \left[ \frac{E_*}{2} e^{i \phi} \right] + \frac{i \theta_d}{2} \frac{1}{\widetilde{r}} e^{i \phi} \left[ \frac{E_*}{2} e^{-i \phi} \right]= \frac{i \theta_d}{2} \frac{1}{\widetilde{r}} E_*.
\end{equation}
It follows directly from Eq.~(\ref{E_z_pm1}) that it is equal to $E_x$ of the circularly polarized beam with $E_0 = E_*$ and $l = -1$, $\sigma = 1$, and $p=0$ that we are using for our comparison. On the other hand, the longitudinal magnetic fields of the two circularly-polarized beams cancel each other out, so that there is no strong $B_x$ close to the axis of the resulting radially polarized beam:
\begin{equation}\label{Rp-Bx}
    B_x = -\frac{\theta_d}{2} \frac{1}{\widetilde{r}} e^{-i \phi} \left[ \frac{E_*}{2} e^{i \phi} \right] + \frac{\theta_d}{2} \frac{1}{\widetilde{r}} e^{i \phi} \left[ \frac{E_*}{2} e^{-i \phi} \right] = 0.
\end{equation}
In contrast to this, $B_x$ of the circularly polarized beam with $l = -1$ and $\sigma = 1$ that we use for our comparison has the same amplitude as $E_x$.

Even though $E_x$ has the same amplitude in the two beams that are being compared, the power in the radially polarized beam is two times lower. In order calculate the power of the radially polarized beam, we note that $B_z = E_y$ and $B_y = -E_z$. The longitudinal component of the Poynting vector is then given by
\begin{equation}
    S_x = \frac{c}{4 \pi} \left[ \left( \mbox{Re} E_y \right)^2 + \left( \mbox{Re} E_z \right)^2 \right] = \frac{c}{4 \pi} E_*^2 \left( \mbox{Re} D \right)^2.
\end{equation}
The peak period-averaged power is
\begin{equation} \label{eq:P_rad}
    P_{rad} = \left\langle \int_0^{2\pi} d \phi \int_{0}^{\infty} S_x r d r \right\rangle = \frac{c w_0^2}{4 \pi}  \int_0^{2\pi} d \phi \int_{0}^{\infty} \left\langle \left( \mbox{Re} D \right)^2 \right\rangle \widetilde{r} d \widetilde{r} 
\end{equation}
with $g(\xi) = 1$. After time-averaging, the expression on the right-hand side is identical to the expression on the right-hand side of \cref{eq:power_poynting_int}. Thus the peak period-averaged power of the radially polarized beam is
\begin{equation} \label{eq:power_rad}
    P_{rad} = \frac{c w_0^2}{8 \pi} E_0^2.
\end{equation}
Not that it is equal to $P_{lin}$, the power of a linearly polarized beam with $E_0 = E_*$, given by \cref{eq:power}. On the other hand, the peak period-averaged power of a circularly polarized beam, $P_{CP}$, with $E_0 = E_*$, $l = -1$, $\sigma = 1$, and $p=0$ is $2P_{lin}$, where $P_{lin}$ is given by \cref{eq:power}. We thus have
\begin{equation}
    P_{CP} = \frac{c w_0^2}{4 \pi} E_0^2 = 2 P_{rad}.
\end{equation}

If both types of beams can be generated at the same power, then a potential advantage of using the radially polarized beam would be its ability to generate higher $E_\parallel$. However, very different optical techniques are employed to generate the two types of beams. The radially polarized beams are produced using transmissive optical elements, which limits the incident power. In contrast to that, the beams with twisted \rc{wavefronts} can be produced by adding an etched mirror~\cite{Longman2020,shi2014} at some point into a traditional laser-system. This method relies of laser reflection, so it does not have the same power limitation. This means that it can potentially be used to generate the desired beams at very high power and reach very strong $E_\parallel$.

To compare the acceleration by the two beams with the same strength of $E_\parallel$, we performed an additional simulation for a 300~TW radially polarized beam. With the exception of the laser power, all other simulation parameters are the same as those listed in 
Table~\ref{table:PIC} and used to obtain the results presented in \cref{Sec-sim}. The electric field structure in the $(x,z)$-plane at $t = -9$~fs is shown in \cref{rp_eb}. The longitudinal electric field structure in the region close to the axis is nearly identical to that shown in
\cref{eb_xz} for the circularly-polarized Laguerre-Gaussian beam. There is no strong $B_\parallel$ near the axis, which agrees with the analytical analysis given by \cref{Rp-Ex} and \cref{Rp-Bx}.

\begin{figure} 
 \centering
 \includegraphics[width=0.85\linewidth]{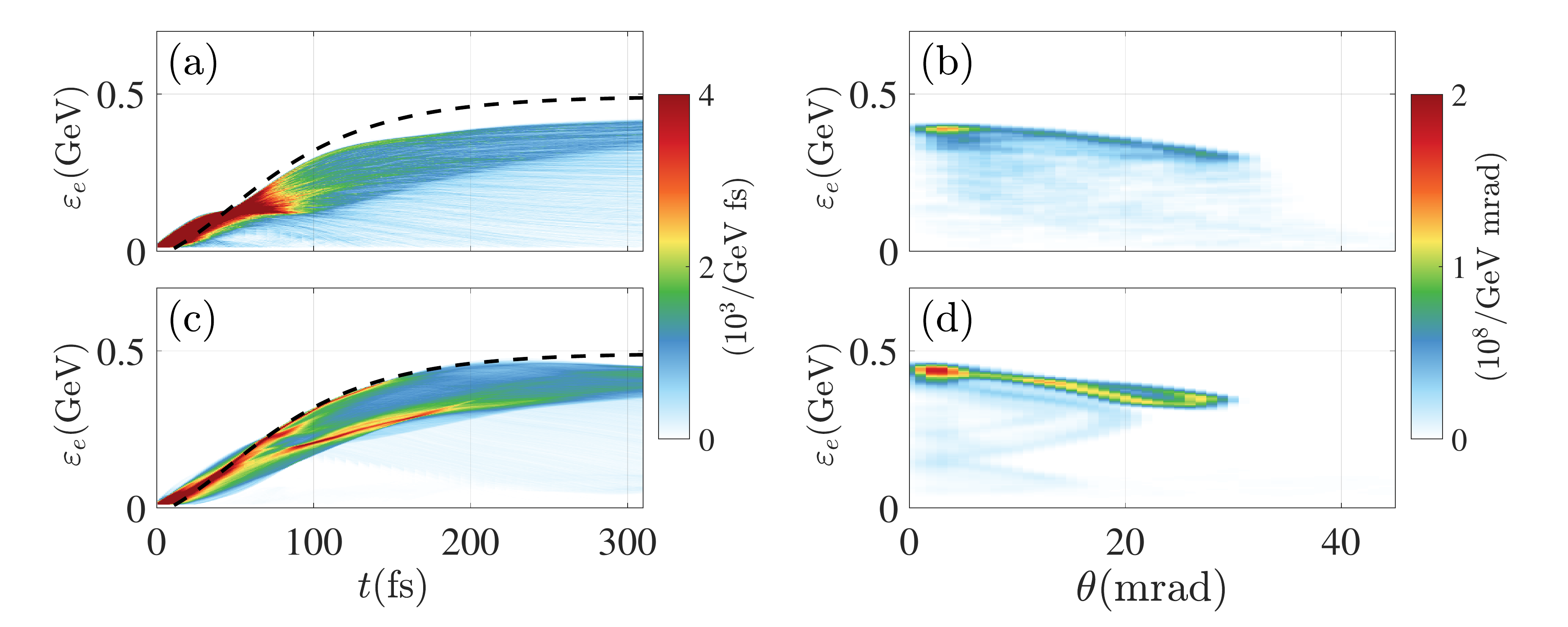}  
\caption{Sensitivity of the electron acceleration by a 300~TW radially polarized beam to simulation parameters. The top row [subplots  (a) and (b)] shows results for the parameters from Table~\ref{table:PIC}. The bottom row [subplots  (c) and (d)] shows results from a simulation with a reduced cell size of 20 nm and a higher-order field solver. The black dashed curve is prediction of \cref{main_result} for \rc{$\Phi_0 = 0.8\pi$}. The left column is the energy distribution as a function of time for the third bunch of electrons. The right column is energy-angle distribution at $t = 261$~fs.} \label{Rp_EngAng2d}
\end{figure}

To obtain a comparable picture of the phenomena in the new simulation, we again focus on the third bunch formed during the reflection process (see Fig~\ref{eDenst_Injct} for the location of this in the circularly-polarized Laguerre-Gaussian case). The evolution of the energy spectrum of this bunch is shown for the radial polarized case in \cref{Rp_EngAng2d}(a) whereas the energy-angle distribution is shown in  \cref{Rp_EngAng2d}(b). These plots can be compared to similar plots in \cref{ebunch}. In the radially polarized case, we see a slightly lower peak kinetic energy at later times. In addition to this, the bunch has a wider energy spectrum. When looking at the energy-angle distribution, the bunch appears to be spread over a slightly wider divergence angle. The  spectra in the radially-polarized case do not show the highly \rc{narrow energy spread} features present in the circularly-polarized case.

Previous studies~\cite{Zaim2020} have shown that the acceleration by radially polarized beams can be affected by the high-harmonic radiation emitted during the reflection. To test for this sensitivity, we performed two more simulations, one for each beam scenario. In these simulations, the cell size is reduced from 25~nm to 20~nm and the default second order Yee scheme is changed to the fourth order version. The results for the circularly-polarized Laguerre-Gaussian beam are unchanged, which is unsurprising given the extensive resolution tests previously run~\cite{Shi2021, Smilei2018}. The new results for the radially-polarized case are visibly different compared to those from the original simulation: there are differences in the maximum energy, energy spread, and the energy-angle distribution. There is also some evidence, which is not plotted here, of a second harmonic forming on reflection of the beam in the radially polarized case. No such high-harmonic generation is evident in the circularly-polarized Laguerre-Gaussian case. The studies focused on radially polarized beams typically use the field solvers specifically designed for accurate propagation of high-harmonic generation~\cite{Zaim2017, Zaim2020,WarpX}. While the results of \cref{Rp_EngAng2d} primarily show the sensitivity of the radially polarized mechanism to \textit{resolving} the high-harmonic generation, there is experimental evidence~\cite{Zaim2020} that high-harmonics do have a significant impact when reflecting this type of beam at an angle away from normal off a plasma mirror.

%\newpage
%\subsection*{References}
%\bibliography{bb}
\bibliographystyle{ieeetr}

\end{document}